\documentclass[fleqn,usenatbib]{mnras}

\usepackage[T1]{fontenc}

\DeclareRobustCommand{\VAN}[3]{#2}
\let\VANthebibliography\thebibliography
\def\thebibliography{\DeclareRobustCommand{\VAN}[3]{##3}\VANthebibliography}

\usepackage{graphicx}	
\graphicspath{ {plots/} }
\usepackage{amsmath}	
\usepackage{amssymb}	
\usepackage{hyperref}
\usepackage{xcolor}

\usepackage{newtxtext,newtxmath}

\title[AGN variability: oscillation or eruption]{Variability In A Low-Mass AGN: Oscillation Or Eruption?}

\author[R. Webbe et al.]{
Robbie Webbe,$^{1}$\thanks{\url{https://orcid.org/0000-0003-1689-3723}}
and A. J. Young$^{1}$\thanks{\url{https://orcid.org/0000-0003-3626-9151}}
\\
$^{1}$H. H. Wills Physics Laboratory, Tyndall Avenue, Bristol, BS8 1TL, UK
}

\date{Accepted XXX. Received YYY; in original form ZZZ}

\pubyear{2022}

\begin{document}
\label{firstpage}
\pagerange{\pageref{firstpage}--\pageref{lastpage}}
\maketitle

\begin{abstract}
Following the discovery of a new class of X-ray variability seen in four galaxies, dubbed Quasi-Periodic Eruptions (QPEs), we reconsider the variability seen in the low-mass AGN 2XMM J123103.2+110648 to ascertain whether it should be considered the fifth QPE host galaxy. We apply the autocorrelation function to two archival \emph{XMM-Newton} observations to determine characteristic timescales for variability of $\sim 13.52$\,ks and $\sim 14.35$\,ks. The modelling of lightcurves, both folded at these timescales and unfolded, indicates that a Gaussian model is preferable over a sinusoidal model, with average durations for the bright phases of 6.17 ks and 7.69 ks. In a broad 0.2--1.0~keV band the average amplitude of the bright phases was found to be 2.86 and 8.56 times the quiescent count rate. The pattern of variability seen in 2XMM J123103.2+110648 cannot be definitively declared as a series of Quasi-Periodic Eruptions. Instead, this suggests there may be a continuum of quasi-periodic variability ranging from eruptions to oscillations being caused by a single mechanism. This offers the possibility of finding further sources that continue to bridge the gap between QPEs and Quasi-Periodic Oscillations. A targeted analysis of 47 observations of 11 other low-mass AGN ($\log(M_\text{BH}) \lesssim 6$) found no evidence of QPE or QPO-like behaviour in a sample of other similar mass objects.

\end{abstract}




\section{Introduction}
\label{sec:intro}

Variability in the X-ray brightness of Active Galactic Nuclei (AGN) has been observed over many decades. These changes in luminosity are well documented, even if not conclusively explained \citep[e.g.,][]{Mushotzky1993,Ulrich1997}. Much of the analysis of AGN X-ray variability in the time-domain has used the cross-correlation as opposed to the autocorrelation function, often comparing X-ray luminosity with emissions in other wavelengths \citep[e.g.,][]{Edelson2002,McHardy2004,Edelson2019} or has been conducted in frequency space by means of power spectra and the Lomb-Scargle periodogram \citep[e.g.,][]{Lin2013,Terashima2012,Gupta2018}. This is typically due to the long timescales over which significant variability occurs in AGN, often over periods of weeks or months, meaning that the variability of interest usually cannot be adequately described within single observations. For this reason the autocorrelation function has instead been used in analysis of the observations of X-ray binaries, which tend to exhibit variability over considerably shorter timescales \citep[e.g.,][]{Gandhi2008,Fukumura2010}. The autocorrelation function is sometimes used in studies of AGN variability to define the minimum timescales over which variability may be un-correlated \citep[e.g.,][]{Gliozzi2002}.

The presence of periodic changes in X-ray luminosity, called Quasi-Periodic Oscillations (QPOs), have been identified in a number of AGN \citep[e.g.,][]{Vaughan2010,Reis2012,Gupta2018} using both time-domain (cross-correlation function) and frequency domain (Power Spectra, Lomb-Scargle periodogram) techniques, although not yet conclusively by examination of lightcurves alone. In the last few years a pattern of variability, being characterised by large changes in luminosity over relatively short timescales and occurring nearly periodically, has been observed in two relatively low mass AGN, GSN 069 \citep{Miniutti2019} and RX J1301.9+2747 \citep{Giustini2020}, and two galaxies which had previously been in quiescent states, 2MASS 02314715-1020112 and 2MASX J02344872-4419325 \citep{Arcodia2021}.  This rapid and large-amplitude variabillity, with increases in luminosity of almost two orders of magnitude over timescales of hours, has been dubbed a Quasi-Periodic Eruption (QPE).

Several models have been proposed to explain this new phenomenon including: Roche lobe overflow from a star on a circular extreme mass ratio inspiral (EMRI) \citep{Arcodia2021,Metzger2022}; accretion from a low-mass white dwarf and partial tidal disruption \citep{King2020}; instabilities in the accretion flow \citep{Miniutti2019}; collisions between an orbiting star and an accretion disk \citep{Xian2021}; gravitational lensing from a binary supermassive black hole system \citep{Ingram2021}; tearing in warped accretion discs caused by Lense-Thirring precession \citep{Raj2021}. Tidal disruption events \citep{Reis2012} and tearing in accretion disks \citep{Musoke2022} have also been proposed as mechanisms for QPOs. Ultimately, in order to better understand this phenomenon we need to increase the number of eruptions seen in known QPE host galaxies, and to increase the number of galaxies which are known to exhibit this behaviour.

The two AGN in which QPEs were first identified share a number of unusual characteristics \citep{Miniutti2019,Giustini2020}: relatively low black hole mass, with $\log(M_\text{BH} / M_\odot) \lesssim 6.5$; higher Eddington ratio; the absence of broad emission lines in optical spectra; ultra-soft X-ray spectrum with a near negligible hard X-ray power law component. The AGN 2XMM J123103.2+110648 (hereafter 2XJ1231) was identified by \cite{Miniutti2019} as sharing many of these characteristics with GSN 069 and RX J1301.9+2747 and was to be considered a candidate AGN for hosting QPEs.

\subsection{2XMM J123103.2+110648}

The AGN 2XJ1231 was first identified by \cite{Lin2012} and \cite{Terashima2012} serendipitously during an \emph{XMM-Newton} observation of the quasar LBQS 1228+1116. It was identified as having a very soft thermal spectrum with a blackbody temperature ranging from 0.12-0.14 keV \citep{Lin2014} to 0.13-0.15 keV \citep{Terashima2012}. At the time, the ratio between the detection of soft and hard X-rays was the greatest detected in AGN, with no detection of X-rays at energies above 2 keV \citep{Terashima2012}. The presence of a QPO was detected by a blind search over a range of folding timescales against a model showing no variability, at a period of 13.71 ks by \cite{Lin2013} and a period of 14.0 ks by \cite{Terashima2012}, with the QPO having a very high fractional RMS variability of up to 50$\%$ in the 1-2 keV energy band \citep{Lin2013}. The black hole mass has been estimated as $M_\text{BH} = 2.8 \times 10^4 M_\odot$ by \cite{Terashima2012}, or in the range $M_\text{BH} = 6.8-10.0\times10^4 M_\odot$ by \cite{Ho2012} with an associated Eddington ratio of between 0.45 and 0.66. In the time between the observations when it was originally discovered in 2003 and follow-up \emph{Swift}/\emph{Chandra} observations between 2013--16 the X-ray flux has decreased by an order of magnitude \citep{Lin2017}.

\bigskip
In this paper we aim to determine, quantitatively, whether the variability seen in X-ray observations of 2XJ1231 is better described by a quasi-periodic oscillation or a series of quasi-periodic eruptions. We do this by comparing lightcurves, and folded lightcurves, with mathematical models designed to represent such behaviour, rather than by analysing products of these lightcurves like Fourier power spectra. We describe the selection and preparation of the data as well as the models against which it will be compared in Section \ref{sec:methods}. In Section \ref{subsec:autocorr} we look at the uses of the autocorrelation function in determining characteristic time periods for variability, and in Section \ref{subsec:modelling} we examine the validity of the models fit to lightcurves. In Sections \ref{sec:discuss} and \ref{sec:conc} we discuss what our results mean for the variability of 2XJ1231, the implications for a wider understanding of quasi-periodic eruptions, and possible implications for the understanding of QPOs.

\begin{figure*}
	\includegraphics[width=\textwidth]{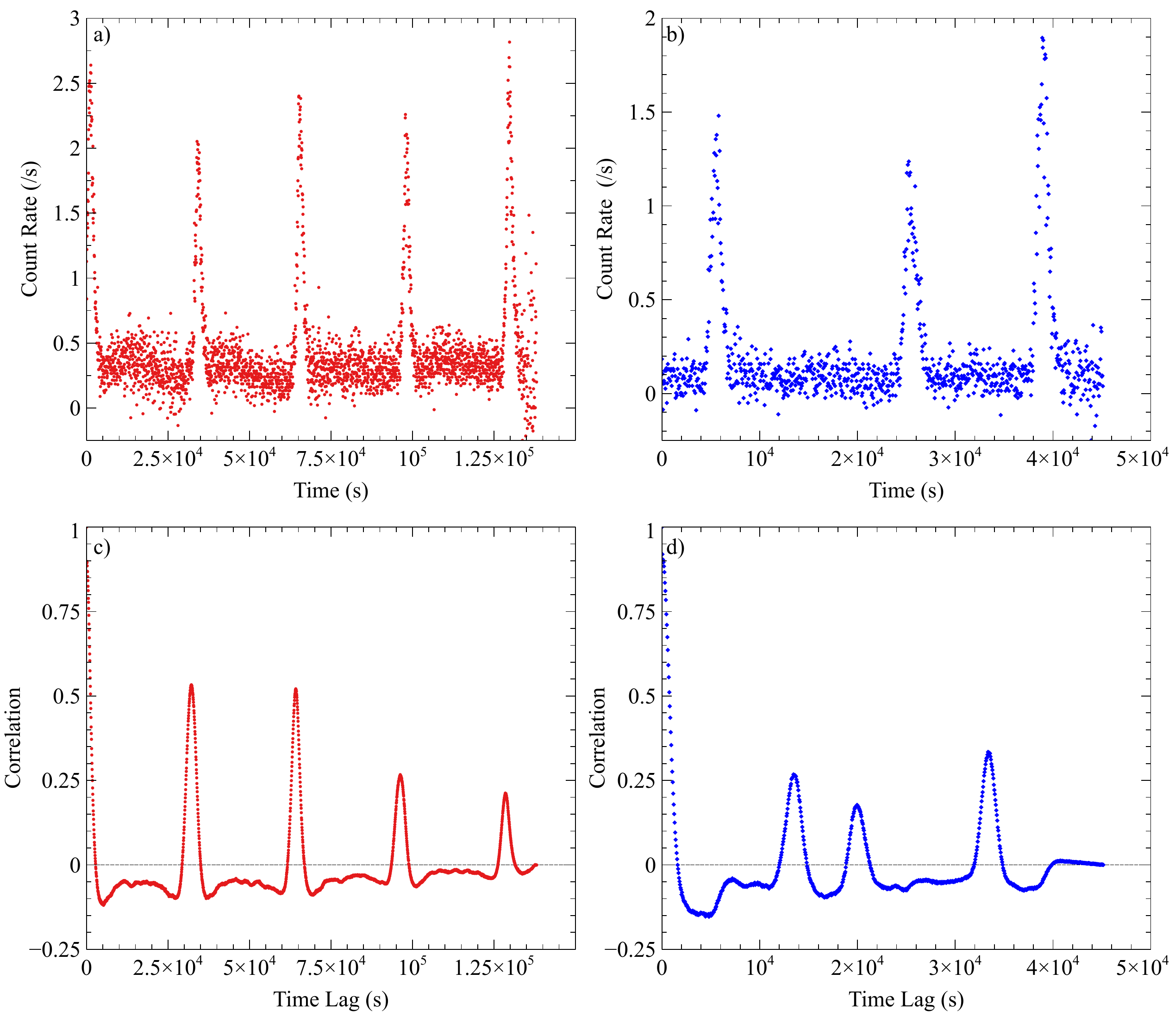}
    \caption{Lightcurves and associated autocorrelation plots for observations of GSN 069 and RX J1301.9+2747. a) Lightcurve of \emph{XMM} observation 0831790701 of GSN 069, for events in the energy range 0.2--2.0 keV binned at 50s intervals. b) Lightcurve of \emph{XMM} observation 0851180501 of RX J1301.9+2747, for events in the energy range 0.2--2.0 keV binned at 50s intervals. c) Autocorrelation function plot for lightcurve in panel 1a). d) Autocorrelation function plot for lightcurve in panel 1b).}
    \label{fig:ac_example}
\end{figure*}

\section{Methods}
\label{sec:methods}

\subsection{Data Reduction}
The analysis described in Section \ref{sec:res} was performed on a pair of \emph{XMM-Newton} observations of 2XJ1231 obtained with the European Photon Imaging Camera (EPIC) pn and Metal Oxide Semi-conductor (MOS) instruments. The observations were taken in imaging mode, and details of the observations are given in Table \ref{tab:obs}. In both observations the source was detected at $6.995\arcmin$ off-axis.

\begin{table}
	\centering
	\caption{Details of \emph{XMM-Newton} observations of 2XJ1231 used in this analysis.}
	\label{tab:obs}
	\begin{tabular}{lcccr} 
		\hline
		OBSID & Label & Start Date & Mode & Exposure (ks) \\
		\hline
		0306630101 & O1 & 13-12-2005 & Imaging & 80.9 \\
		0306630201 & O2 & 17-12-2005 & Imaging & 99.5 \\
		\hline
	\end{tabular}
\end{table}

The observations were reprocessed using XMMSAS\footnote{version xmmsas$\_$20190531$\_$1155-18.0.0}. Events for the source were filtered as PI in [200:2000] to give events with energies in the range from 0.2-2.0 keV for parity with analysis of other QPE host galaxies. The source and background respectively were extracted using a circle, centred on the source, with radius $17.5\arcsec$, and an annulus centred at the same point with inner and outer radii of $25\arcsec$ and $37\farcs5$. The lightcurves for the pn and MOS2 instruments were then barycenter corrected and background subtracted using \texttt{lccorr} to produce the curves presented for analysis in this work. Background flares were detected at the end of both observations and the sections containing them were removed, leaving total good exposure times of 63.35 ks and 89.82 ks for observations O1 and O2 respectively. For both observations we used the photon data from the EPIC pn and MOS2 detectors to create combined lightcurves in order to improve the signal to noise ratio. Events from the MOS1 detector were not used as the section of the detector which would have contained the off-axis source had no available data. The combined pn/MOS2 lightcurves were then binned at a rate of 10 seconds.

\subsection{Autocorrelation}
\label{sec:ac} 
The autocorrelation function calculates the product of the deviance from the mean for pairs of points at a given separation in time. For the lightcurves presented here the normalised autocorrelation function (ACF) as a function of time lag $\tau$ is realised as
\begin{eqnarray}
    ACF(\tau)=  \frac{1}{ACF(0)}
    \sum_{i=1}^{N}
    (x(t_i) - \overline{x}) (x(t_i + \tau) - \overline{x}) \\
     \text{where } N = \left\lfloor \frac{T-\tau}{\Delta t} \right\rfloor , \nonumber
	\label{eq:acf}
\end{eqnarray}
where $\Delta t$ is the bin size, the factor $ACF(0)$ normalises the autocorrelation to 1 at a time lag of 0s, T is the full length of the lightcurve, $t_i$ are the times of the bins in the lightcurve, $x(t_i)$ are the count rates at those times and $\overline{x}$ is the average of the lightcurve. If a pair of points are correlated then their product will be positive, and if anti-correlated the product will be negative. Thus, for a lightcurve, time lags which are strongly correlated or anti-correlated will have very high or low values for the autocorrelation function. The autocorrelation functions were created using the \texttt{AutoCorrelation} class within the python module \texttt{stingray} version 0.3 \citep{Huppenkothen2019}. The ACF can then be examined for peaks to determine time lags for which the lightcurve is closely correlated. Figure \ref{fig:ac_example} shows examples of the lightcurves and associated autocorrelation plots for \emph{XMM} observation 0831790701 of GSN 069 and observation 0851180501 of RX J1301.9+2747 both of which are known to contain QPEs. The autocorrelation function for lightcurves containing QPEs display clear peaks at the recurrence time between eruptions ($\sim$32 ks), and multiples of that time (64, 96 \& 128 ks) for the observation of GSN 069. The three peaks in the autocorrelation function of the observation of RX J1301.9+2747 show the more complex combination of long and short recurrence times, with peaks at $\sim$13, 20 and 33 ks. Fitting of models to the Autocorrelation curves is performed using the \texttt{curve$\_$fit} function within the \texttt{scipy} module for python. Close to $\tau = 0$ the ACF will be very large, as points at small time lags should be closely correlated in the absence of very fast large scale variability, and will then decay to zero. For a periodic process there will then be a series of peaks at the characteristic timescale of variability and at multiples of this time. In order to provide the best quality folded lightcurves for fitting we will use the first peak from the ACF to determine the minimum period for the quasi-periodic oscillation or eruption.

\subsection{Models}
\label{sec:models}

In this analysis we will be fitting two different models to the lightcurves. In order to model the presence of a quasi-periodic oscillation in the lightcurve we will use a sinusoidal wave (as used to model QPOs in, e.g., \citeauthor{Song2020} 2020), and to model the presence of quasi-periodic eruptions we will use Gaussian curves \citep{Miniutti2019,Giustini2020,Arcodia2021}. The waveform which will be used to model sinusoidal behaviour will be of the form
\begin{equation}
    x(t)=  (x_\text{q} + mt) + A \sin\left(\frac{2 \pi (t-t_0)}{t_\text{rec}}\right)  
	\label{eq:wave}
\end{equation}
where $x_\text{q}$ is related to the median level between periods of high and low luminosity, $m$ is the rate of change of the quiescent count rate, $A$ defines the amplitude of the oscillations, $t_0$ defines the offset of the oscillations with regards to our observing window, and $t_\text{rec}$ is the recurrence timescale. The bright phases will be modelled using a series of Gaussian curves giving a lightcurve the form
\begin{equation}
    x(t)=  (x_\text{q} + mt) + \sum_{i=1}^{N} A_i \exp \left( \frac{-\ln{(2)}(t-t_{\text{peak},i})^2}{t_{\text{dur},i}^2} \right)
	\label{eq:ngauss}
\end{equation}
where $N$ is the number of bright phases seen in the lightcurve, $x_\text{q}$ is the quiescent count rate, $m$ is the rate of change of the quiescent count rate, $A_i$ is the amplitude of the $i^\text{th}$ bright phase, $t_{\text{peak},i}$ is the peak time of the $i^\text{th}$ bright phase during the observation, and $t_{\text{dur},i}$ is the duration of the $i^\text{th}$ bright phase, defined as the FWHM of the Gaussian curve. In order to consider a variety of these models we will fit the Gaussian model both with and without a fixed amplitude, recurrence time and duration across the bright phases seen in the lightcurves, and will allow a linearly increasing or decreasing component for the baseline levels in both the sinusoidal and Gaussian models. 

For fitting the folded lightcurves we will use variations of the models presented above with the recurrence time as determined by the autocorrelation function. To allow for multiple bright phases affecting the folding window in the Gaussian model we will constrain the number of bright phases to be three, such that a window is expected to contain one bright phase which may be affected by the tails of those on either side. No such correction is needed for the sinusoidal model as the profile of the variability will not be affected by the phase. The best fitting models are reported in section \ref{subsec:modelling}. In order to perform the fitting itself we will use the \texttt{GaussianLoglikelihood} and \texttt{ParameterEstimation} classes contained in the \texttt{stingray} \citep{Huppenkothen2019} module for python.

\begin{table*}
	\centering
	\caption{Parameters and goodness of fit statistic for model fits to full lightcurves of observations O1 and O2 binned at a rate of 250s. The Gaussian model was fit with 5 bright phases for observation O1, and 6 bright phases for observation O2 when Gaussian features are fit, and 7 bright phases when allowed to vary. For those models where the Gaussian height, width and recurrence times were allowed to vary the parameters are the average and standard deviation of the best fitting amplitudes ($A$), durations ($t_\text{dur}$) and recurrence times ($t_\text{rec}$) of the peaks. Parameters are as described in section \ref{sec:models}.}
	\label{tab:lc_fits}
	\begin{tabular}{lcccccccr} %
		\hline
		Observation & Model & $x_\text{q}$ ($\text{s}^{-1}$) & m ( $\times 10^{-7} \text{s}^{-2}$)& $A$ ($\text{s}^{-1}$) & $t_\text{dur}$ (s)& $t_\text{rec}$ (s) & dof & $\chi^2_\nu$ \\
		\hline
		O1 & Wave (QPO) + Const & $0.103 \pm 0.001$ & -- & $0.051\pm0.002$ & -- & $13330\pm10$ & 249 & 5.20\\
		O1 & Wave + Lin & $0.097\pm0.002$ & $1.92\pm0.54$ & $0.050\pm0.001$ & -- & $13370\pm40$ & 248 & 5.17\\
		O1 & Gauss (QPE) + Const & $0.058\pm0.017$ & -- & $0.109\pm0.003$ & $5190\pm1440$ & $13320\pm20$ & 248 & 4.96 \\
		O1 & Gauss + Linear & $0.053\pm0.007$ & $1.86\pm0.93$ & $0.108\pm0.004$ & $5170\pm760$ & $13340\pm20$ & 247 & 4.93 \\
		O1 & Variable Gauss + Const & $0.048\pm0.001$ & -- & $0.125\pm0.024$ & $6170\pm1000$ & $13700\pm1400$ & 237 & 3.79\\
		O1 & Variable Gauss + Lin & $0.052\pm0.003$ & $-1.88\pm0.72$ & $0.128\pm0.029$ & $6420\pm1240$ & $13800\pm1400$ & 236 & 3.79\\
		O2 & Wave + Const & $0.0713\pm0.001$ & -- & $0.034\pm0.001$ & -- & $13980\pm20$ & 355 & 6.75\\
		O2 & Wave + Lin & $0.074\pm0.001$ & $-0.79 \pm 0.24$ & $0.034\pm0.001$ & -- & $13960\pm30$ & 354 & 6.74\\
		O2 & Gauss + Const & $0.025\pm0.002$ & -- & $0.087\pm0.002$ & $7910\pm70$ & $13500\pm20$ & 354 & 5.52\\
		O2 & Gauss + Lin & $0.010\pm0.003$ & $2.50\pm0.29$ & $0.092\pm0.003$ & $7988\pm1$ & $13492\pm1$ & 353 & 5.23\\
		O2 & Variable Gauss + Const & $0.009\pm0.012$ & -- & $0.103\pm0.016$ & $8350\pm1480$ & $14000\pm900$ & 337 & 4.15 \\
		O2 & Variable Gauss + Lin & $0.022\pm0.002$ & $-2.55\pm0.76$ & $0.102\pm0.017$ & $8270\pm1610$ & $14000\pm800$ & 336 & 4.15 \\
		\hline
	\end{tabular}
\end{table*}

\section{Results}
\label{sec:res}

\subsection{Lightcurve Modelling}
\label{subsec:modelling}

The observations O1 and O2 were fitted to each of the models described in section \ref{sec:models}. To avoid empty time bins in the lightcurves, for the purpose of fitting, the lightcurves were rebinned at a rate of 250s. The best fitting models to the lightcurves of each observation, were determined by minimising the reduced chi-squared statistic as defined as
\begin{equation}
    \chi^2_\nu = \frac{1}{n_{\text{dof}}} \sum_{i} \frac{(O_i - E_i)^2}{\sigma^2_i}
	\label{eq:red_chi}
\end{equation}
where $O_i$ are the observed values of the count rate, $E_i$ are the expected values for the rate as defined by the model being fit, $n_{\text{dof}}$ is the number of degrees of freedom of the model, and $\sigma_i$ are the uncertainties on the count rate as calculated by \texttt{stingray} as frequentist central confidence intervals.

The best fitting models to the full lightcurves of O1 and O2 were both of a series of Gaussian bright phases where the bright phase widths, heights and recurrence times were allowed to vary. For O1 the quiescent countrate was linearly decreasing, whereas it was constant for O2. The best fitting Gaussian models to O1 used five peaks, whereas the best fitting Gaussian models to O2 used 6 peaks when the features of the bright phases were fixed and 7 peaks when allowed to vary. This difference is due to the position of the 7$^\text{th}$ peak at the very end of the observation. For the sinusoidal and fixed Gaussian model there was a clear distinction in the recurrence times across O1 and O2, with the recurrence time for O1 being around 13.3 ks as opposed to 14.0 ks or 13.5 ks for O2 in the sinusoidal and Gaussian models. In the best fitting models when the characteristics of the Gaussians were allowed to vary the amplitudes ranged from 0.053--0.153 counts/s and the durations of the bright phases ranged from 3.81--6.84 ks for O1. The amplitudes ranged from 0.061--0.106 counts/s and the durations of the bright phases ranged from 5.18--10.04 ks for O2. When the profiles of the Gaussians were allowed to vary they were significantly broader, of a greater amplitude, and with longer recurrence times for O1. The full details of the parameters for the best fitting models are given in Table \ref{tab:lc_fits}.

In both observations there was a hierarchy as to the goodness of fit, with the unconstrained Gaussian model being most suitable and the wave model providing the worst fit to the lightcurves. The fits to the full lightcurves indicated that in the case of all models the baseline count rates were lower for the second observation than for the first. 

\begin{figure}[h]
\includegraphics[width=\linewidth]{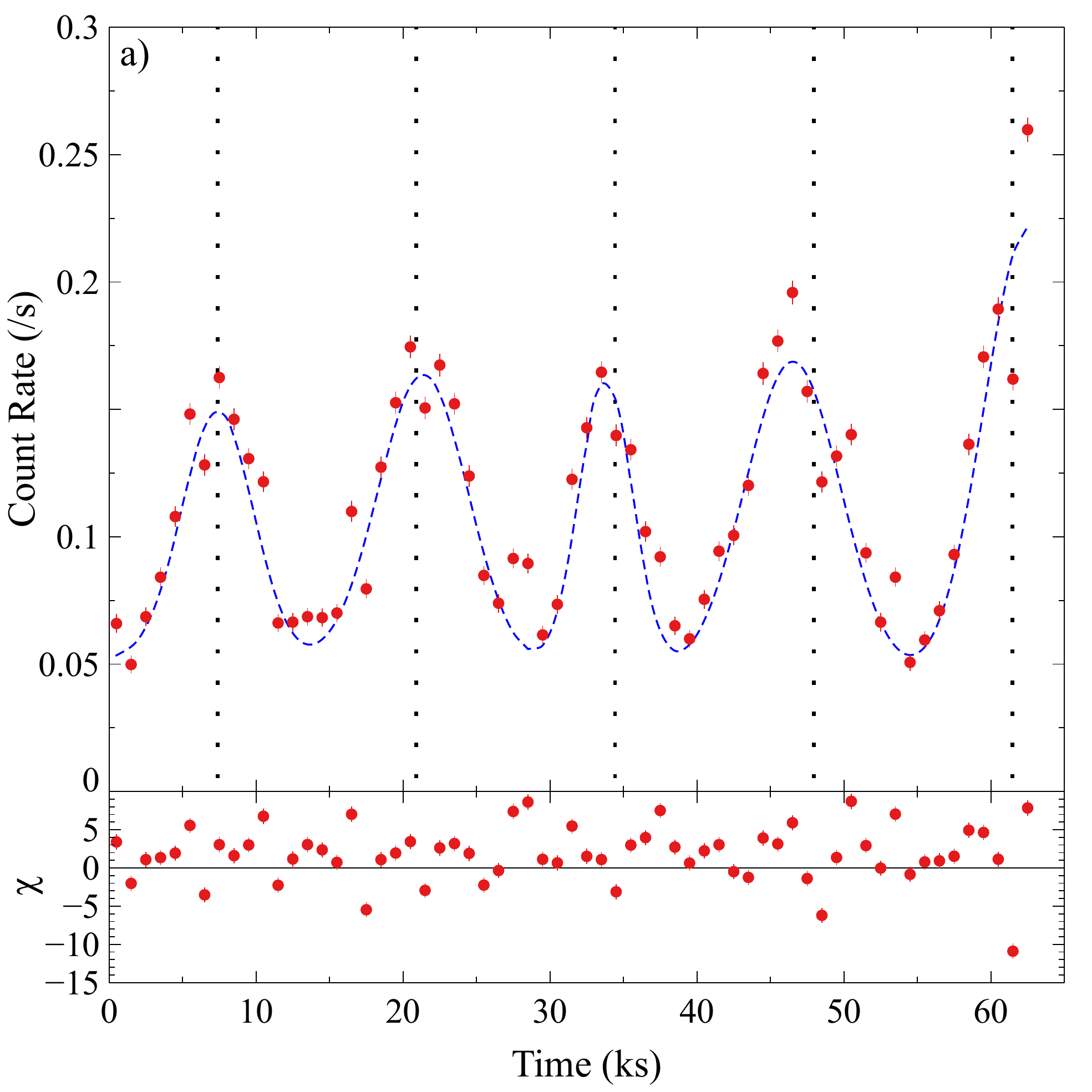}
\includegraphics[width=\linewidth]{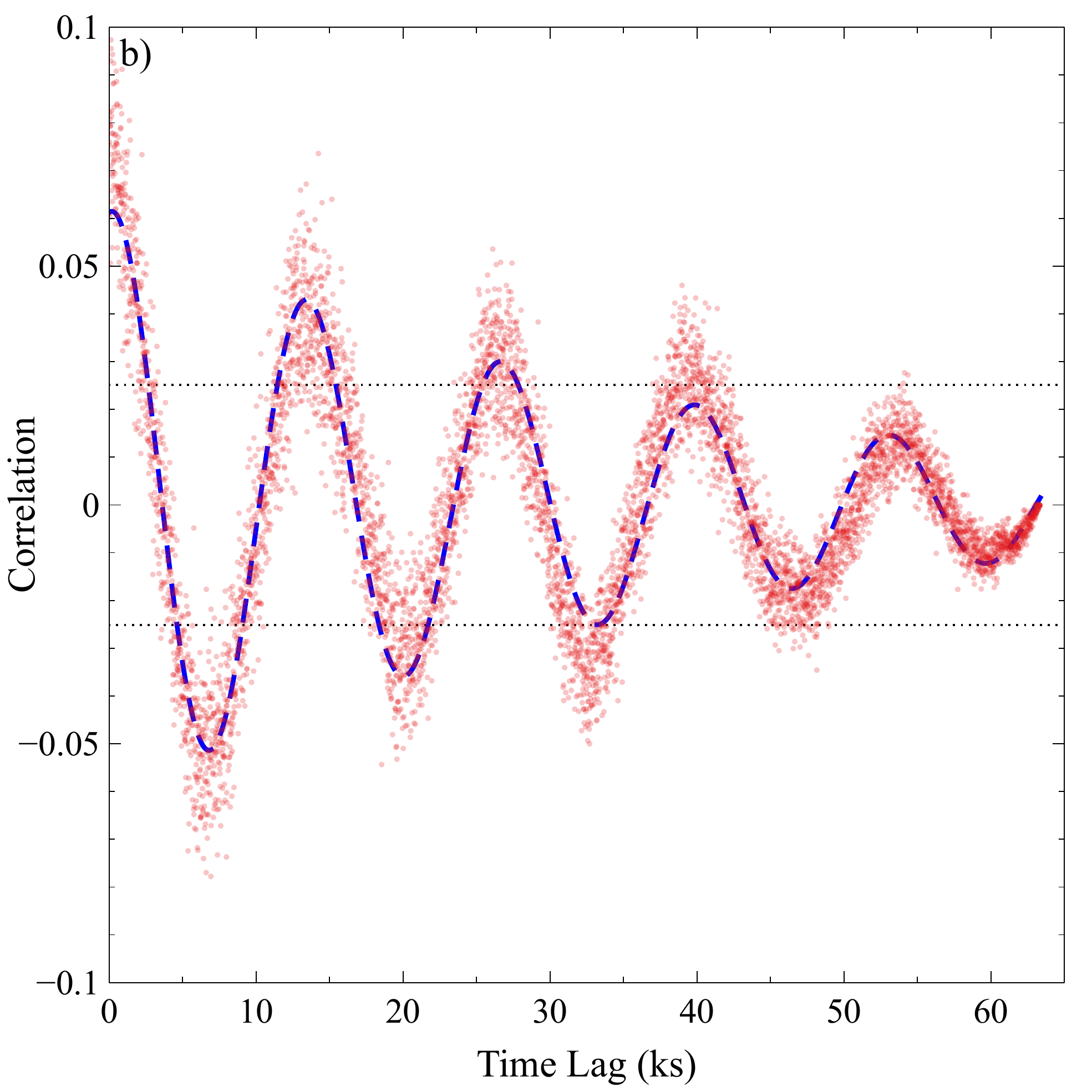}
\caption{Variable Gaussian model fit and autocorrelation function plot for unfolded lightcurve of observation O1 of the AGN 2XJ1231. a) Gaussian fit to lightcurve of observation O1, with a linearly decreasing baseline count rate. Curve is binned at 1000s for clarity and the lower panel shows the deviation of each point from the model. Dashed line shows the best fitting model, and vertical dotted lines show the position of the first peak, and multiples of the recurrence time, 13.52ks, as per the analysis of the autocorrelation function. b) Fitted autocorrelation plot, with dashed line showing the best fitting exponentially damped wave. The point corresponding to $ACF(\tau=0)=1$ has been omitted to better illustrate the behaviour for $\tau$ greater than 0. Dotted lines represent the 5$\sigma$ confidence level.}
\label{fig:O1_full_lc_fit_and_acf}
\end{figure}

\begin{figure}
\includegraphics[width=\linewidth]{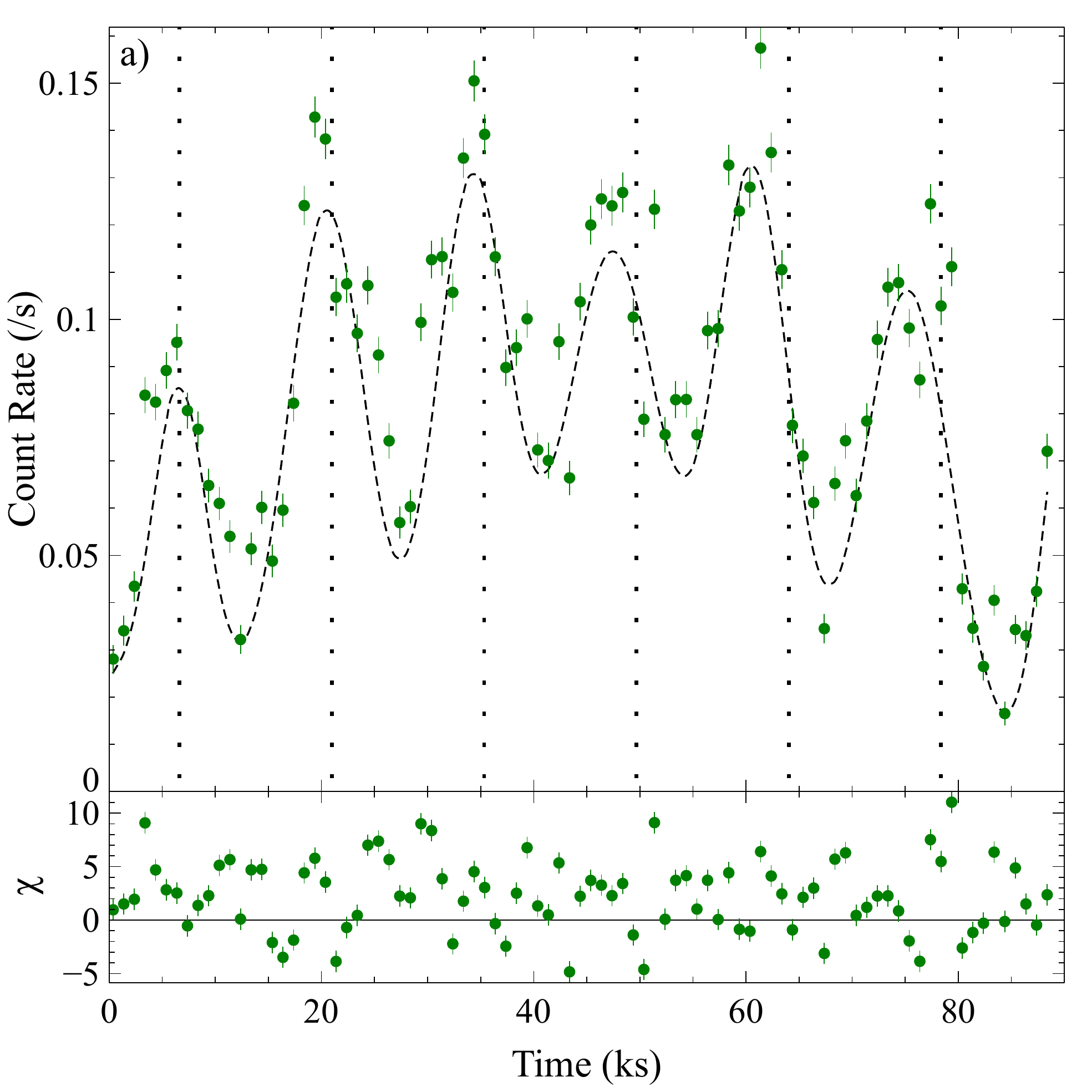}
\includegraphics[width=\linewidth]{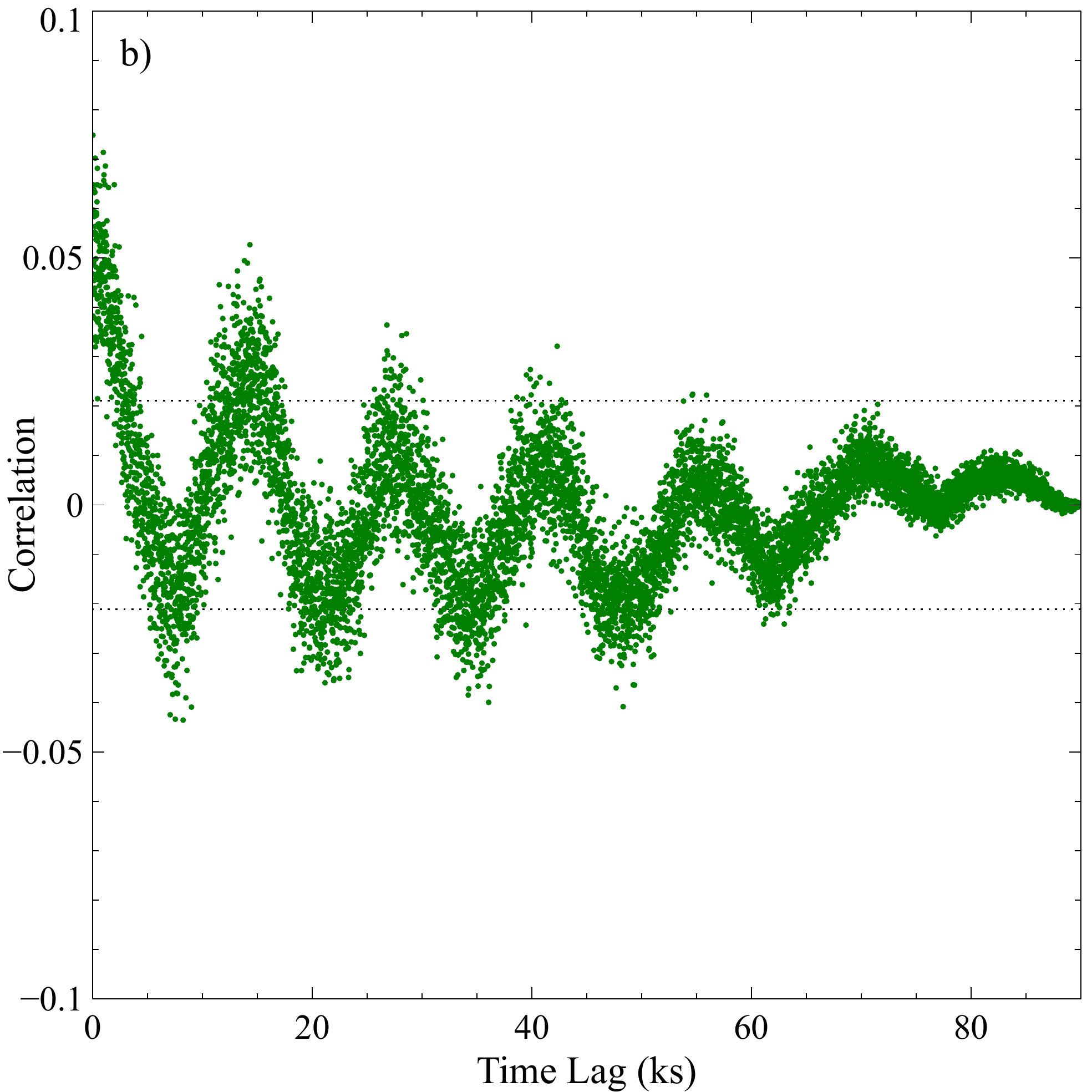}
\caption{Variable Gaussian model fit and autocorrelation function plot for unfolded lightcurve of observation O2 of the AGN 2XJ1231. a) Gaussian fit to lightcurve of observation O2, with a linearly decreasing baseline count rate. Curve is binned at 1000s for clarity and the lower panel shows the deviation of each point from the model. Dashed line shows the best fitting model, and vertical dotted lines show the position of the first peak, and multiples of the recurrence time, 14.35ks, as per the analysis of the autocorrelation function. b) Autocorrelation plot. The point corresponding to $ACF(\tau=0)=1$ has been omitted to better illustrate the behaviour for $\tau$ greater than 0. Dotted lines represent the 5$\sigma$ confidence level.}
\label{fig:O2_full_lc_fit_and_acf}
\end{figure}

\subsection{Autocorrelation Function}
\label{subsec:autocorr}

Following the identification of coherent, apparently quasi-periodic, variability in the lightcurves of the observations we consider the use of the autocorrelation to determine the characteristic timescale of variability. The lightcurves for the observations O1 and O2 were processed as described in section \ref{sec:methods}. Autocorrelation curves were created for both observations using the lightcurves with a time binning of 10s and are shown in figures \ref{fig:O1_full_lc_fit_and_acf} and \ref{fig:O2_full_lc_fit_and_acf}.

Due to the absence of distinctive peaks in the autocorrelation curves, as seen in the autocorrelation curves for observations of other AGN containing QPEs (see Figure \ref{fig:ac_example}), we fit an exponentially damped wave to the autocorrelation function of the form:
\begin{equation}
    ACF(\tau)=  c + A \exp\left(-\frac{\tau}{t_{d}}\right)  \times \cos(2\pi f (\tau - \tau_0))
	\label{eq:edw}
\end{equation}
in order to find the location of the first peak. This exponentially damped wave provides a good description of the autocorrelation function and has been used in other variability studies of AGN \citep{Chen2022}. See Figure \ref{fig:O1_full_lc_fit_and_acf}b) for an example of the fitting to observation O1. The values of $\tau$ at the first peak of this wave were (13.52$\pm$0.32) ks and (14.35$\pm$0.94) ks for observations O1 and O2 respectively, and were used to create folded lightcurves. The characteristic timescales of variablility are, however, significantly different from those determined by model fitting to the raw lightcurves.
 
\subsection{Folded Lightcurves}
\label{subsec:folded}
\begin{figure}
\includegraphics[width=\linewidth]{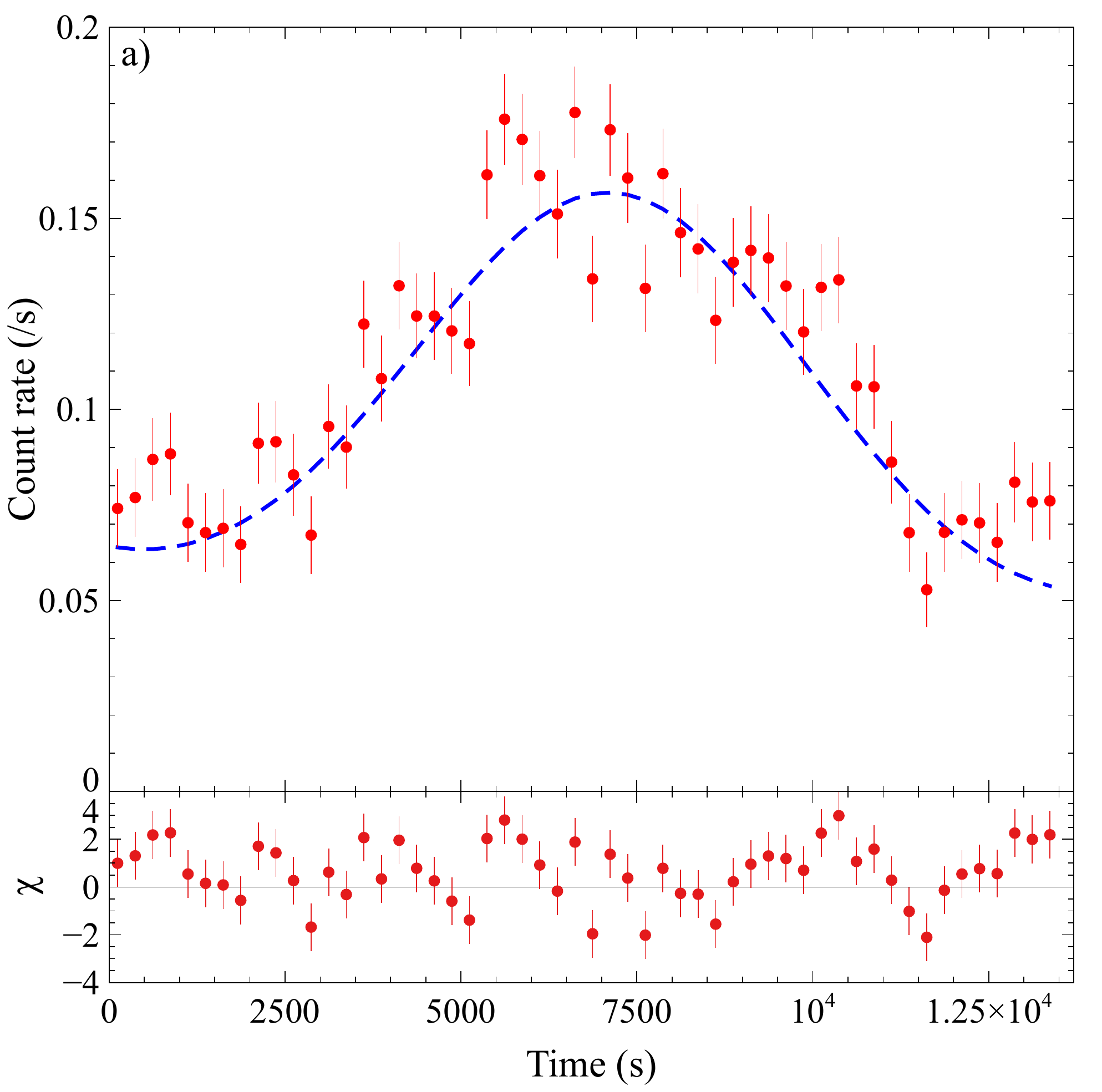}
\includegraphics[width=\linewidth]{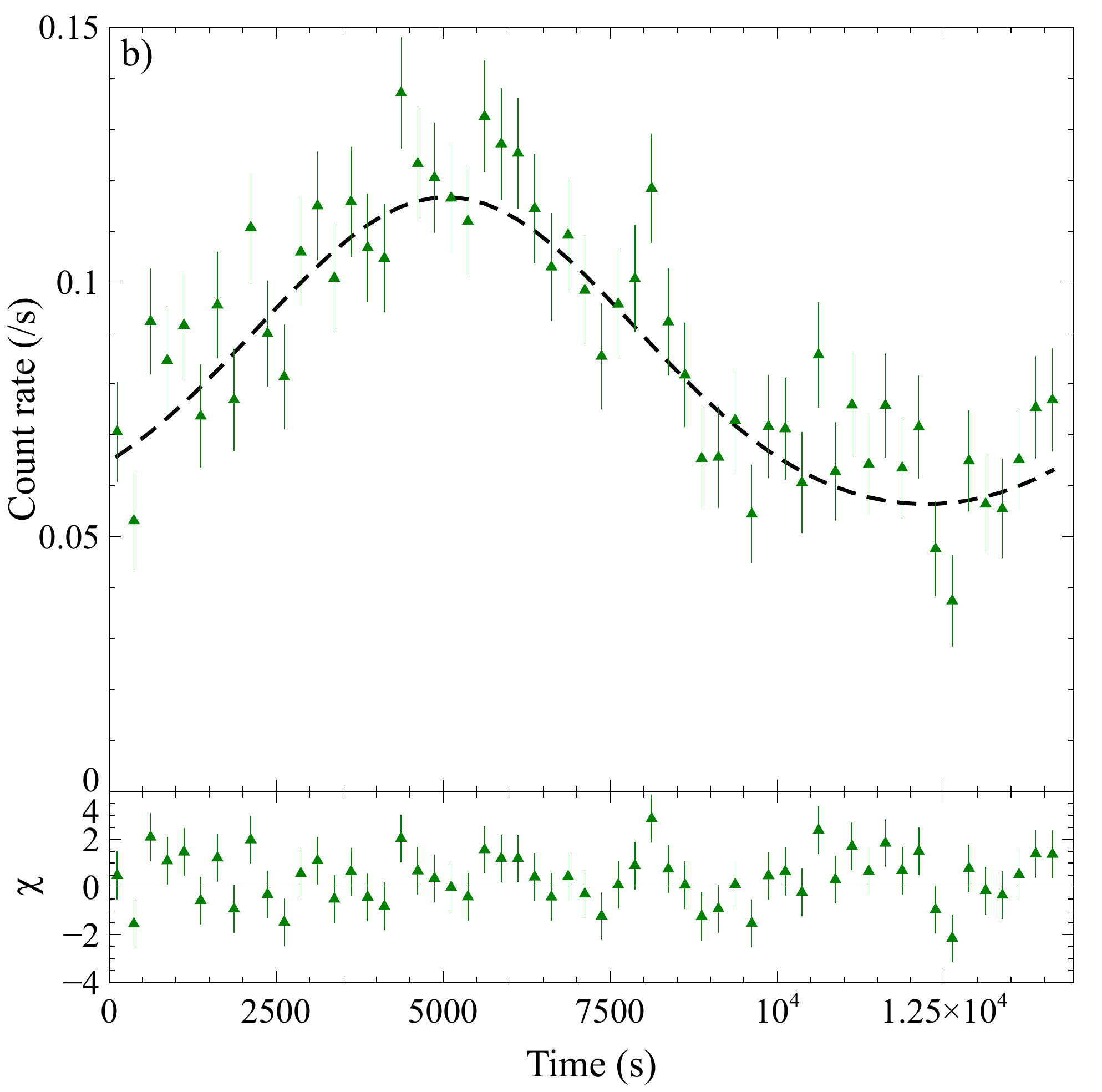}
\caption{Best fits to folded 0.2--2.0 keV lightcurves of the two observations of the AGN 2XJ1231. Figures are shown with lightcurves binned at a rate of 250s for clarity. The lower panels for each fit show the deviation of each point from the model. a) Gaussian fit to lightcurve of observation O1, folded at a period of 13.52 ks with a linearly decreasing baseline count rate. b) Gaussian fit to lightcurve of observation O2, folded at a period of 14.35 ks with a constant quiescent count rate.}
\label{fig:folded_fits}
\end{figure}
\begin{table*}
	\centering
	\caption{Parameters and goodness of fit statistic for model fits to folded lightcurves of observations O1 and O2 of 2XJ1231 binned at a rate of 50s. Folding timescales are 13.52 ks and 14.35 ks for O1 and O2 respectively. Parameters are described in section \ref{sec:models}.}
	\label{tab:folded_fits}
	\begin{tabular}{lcccccccr} %
		\hline
		Observation & Model & $x_\text{q}$ ($\text{s}^{-1}$) & m ( $\times 10^{-7} \text{s}^{-2}$)& $A$ ($\text{s}^{-1}$) & $t_\text{0}$ (s)& $t_\text{dur}$ (s)& dof & $\chi^2_\nu$ \\
		\hline
		O1 & Wave (QPO) + Constant & $0.103 \pm 0.001$ & -- & $0.048 \pm 0.003$ & $6980 \pm 110$ & -- & 267 & 1.290 \\
		O1 & Wave + Linear & $0.111 \pm 0.005$ & $-10.7\pm6.6$ & $0.049 \pm 0.003$ & $7200 \pm 170$ & -- & 266 & 1.285 \\
		O1 & Gaussian (QPE) + Constant & $0.049 \pm 0.004$ & -- & $0.108 \pm 0.006$ & $6990 \pm 80$ & $6390 \pm 210$ & 266 & 1.283 \\
		O1 & Gaussian + Linear & $0.052 \pm 0.037$ & $-8.22 \pm 4.69$ & $0.110 \pm 0.025$ & $7150 \pm 60$ & $6540 \pm 2510$ & 265 & 1.280\\
		O2 & Wave + Const & $0.084 \pm 0.002$ & -- & $0.030 \pm 0.002$ & $5040 \pm 10$ & -- & 283 & 0.866 \\
		O2 & Wave + Lin & $0.079 \pm 0.004$ & $6.75 \pm 6.12$ & $0.032 \pm 0.003$ & $4920 \pm 150$ & -- & 283 & 0.864\\
		O2 & Gauss + Const & $0.051 \pm 0.002$ & -- & $0.066 \pm 0.004$ & $5050 \pm 220$ & $6690 \pm 110$ & 283 & 0.863\\
		O2 & Gauss + Lin & $0.043 \pm 0.007$ & $3.44 \pm 5.47$ & $0.072 \pm 0.007$ & $4980 \pm 50$ & $7200 \pm 30$ & 282 & 0.865\\
		\hline
	\end{tabular}
\end{table*}

The folded lightcurves were rebinned at a rate of 50s for the purposes of avoiding empty bins during fitting. The best fitting model to the folded lightcurve of O1 was of that of Gaussian bright phases with a linearly decreasing baseline count rate, and the best fitting model to the folded lightcurve of O2 was of that of Gaussian bright phases with a constant quiescent count rate. For observation O1 the full amplitudes (peak to trough) of the wave fit were approximately equal to the baseline rate, while for the Gaussian fits the amplitude was approximately double that of the quiescent rate. For observation O2, the amplitude of the wave oscillations was approximately $3/4$ the magnitude of the baseline count rate, while the amplitudes of the Gaussian profiles were of the same order as the quiescent rate. The fits to the folded lightcurves indicated that the FWHM of the Gaussian bright phases was approximately 50$\%$ of the folding timescale. The full details of the parameters for the best fitting models are given in Table \ref{tab:folded_fits}. The quality of fit was very similar across the four models for each of the observations with very little difference in the reduced chi-squared value within the observations. In all cases where the baseline countrate was modelled by a linear trend the slope of the line was consistent with 0 within 2$\sigma$, and in many cases within one. For the Gaussian model the ratio between the quiescent rate and amplitude decreased from O1 to O2 (four days later), while the duration increased. For the full lightcurves the best fitting sinusoidal and Gaussian profiles for observations O1 and O2 gave quiescent rates and amplitudes of variability which were approximately the same as those found for the folded lightcurves with the exception of the Gaussian fits to O2. In contrast with the fits to the full lightcurves, the amplitudes of variability were less for the second observation in the Gaussian models when compared with the baseline rates, and $t_\text{dur}$ was smaller. Across the board the reduced chi-squared values were significantly higher when using the full lightcurve over the folded lightcurves, even for similar numbers of degrees of freedom and with a broader time binning.

\subsection{Energy Spectra}
\label{subsec:energyspec}
To determine differences between the spectra during bright and low phases we extracted events for the two observations using selections in time, as making a cut by count rate would be problematic for O2. We defined events during the peaks of the bright phases as being within 25$\%$ of $t_\text{dur}$ from the peak time for each individual bright phase, and events during the low phases as being more than 75$\%$ of $t_\text{dur}$ from the peak time for each individual bright phase. The spectra were then rebinned to ensure there were at least 5 count in each energy bin. We then fit the spectra for energies between 0.2 and 5.0~keV. Fitting each of these spectra to a {\tt tbabs $\times$ bbody} model in {\tt xspec}, with $N_H = 2.22 \times 10^{20} \text{ cm}^{-2}$ fixed at the Galactic value, indicated that the temperature did not significantly change between O1 and O2, or between the bright and low phases in each observation. The full results are reported in Table \ref{tab:espec}.

\begingroup
\renewcommand{\arraystretch}{1.5}
\begin{table}
	\centering
	\caption{Parameters and goodness of fit statistic for model fits to folded lightcurves of observations O1 and O2 of 2XJ1231 binned at a rate of 50s. Folding timescales are 13.52 ks and 14.35 ks for O1 and O2 respectively. Parameters are described in section \ref{sec:models}.}
	\label{tab:espec}
	\begin{tabular}{lccccr} %
		\hline
		Obs & Phase & Norm ($\times 10^{-6})$ & kT (keV) & $\chi^{2}$ & dof \\
		\hline
		O1 & Bright & $2.59^{+0.19}_{-0.18}$ & $0.125^{+0.006}_{-0.006}$ & 116.17 & 130  \\
		O1 & Low & $0.954^{+0.106}_{-0.103}$ & $0.132^{+0.010}_{-0.010}$ & 67.19 & 80  \\
		O2 & Bright & $1.91^{+0.14}_{-0.13}$ & $0.118^{+0.005}_{-0.005}$ & 121.2 & 132\\
		O2 & Low & $0.937^{+0.138}_{-0.127}$ & $0.112^{+0.010}_{-0.010}$ & 55.45 & 55\\
		\hline
	\end{tabular}
\end{table}
\endgroup

\section{Discussion}
\label{sec:discuss}

The first QPEs were detected in the active galaxy GSN 069 by \citet{Miniutti2019} in one \emph{Chandra} and two \emph{XMM-Newton} observations taken from December 2018 through to February 2019. In the energy range of 0.2--2.0 keV, the QPEs seen in GSN 069 had recurrence times, i.e. the time from the peak of one eruption to the peak of the next, ranging from 29 ks to over 32 ks, and their durations, defined as the full width at half maximum for the best fitting Gaussian curves to the eruptions, ranged between 1.7 ks and 1.9 ks. The size, time and durations of the eruptions did vary when narrower energy bands within this range were also considered. The amplitude of the eruption increased with increasing energy up to the 0.6--0.8~keV band \citep{Miniutti2019}, and then decreased. The durations of eruptions decreased, and the time at which the eruptions peaked was brought earlier, with increasing energy \citep{Miniutti2019}. A previous observation with \emph{XMM} in 2014 had seen no evidence of QPEs in GSN 069.

QPEs were then detected by \citet{Giustini2020} in the active galaxy RX J1301.9+2747 in an archival \emph{XMM} observation from 2000 and in a further \emph{XMM} observation in 2019. The eruptions in this AGN showed many of the same trends as those in GSN 069 including the increasing amplitudes and shorter durations of eruptions in higher energy bands. The eruptions in RX J1301.9+2747 were, however, shorter lived with an average duration of 1.2 ks and they also showed an alternating pattern of larger and smaller amplitude eruptions alongside a pattern of shorter and longer recurrence times between eruptions (20 ks and 13.5 ks).

More recently, QPEs were discovered by \cite{Arcodia2021} in two galaxies whose optical spectra had previously shown no broad line evidence of an accreting black hole. This was achieved by means of a blind search with the \emph{eROSITA} instrument which were then followed up with observations by \emph{XMM-Newton} or \emph{NICER}. The characteristics of the eruptions from the galaxies 2MASS 02314715-1020112 and 2MASX J02344872-4419325 were inconsistent with the predictions from radiation-pressure-driven instability models based on the eruptions seen from the two AGN and widened the temporal parameters of observed QPEs. A further object, XMMSL1 J024916.6-041244, was determined by \citet{Chakraborty2021} to exhibit one full eruption and possibly the start of a second in an archival \emph{XMM} observation from 2006. There was no such activity in a follow-up observation from August 2021, and as such the quasi-periodicity of any eruptions is unclear.
\begin{figure}
	\includegraphics[width=\linewidth]{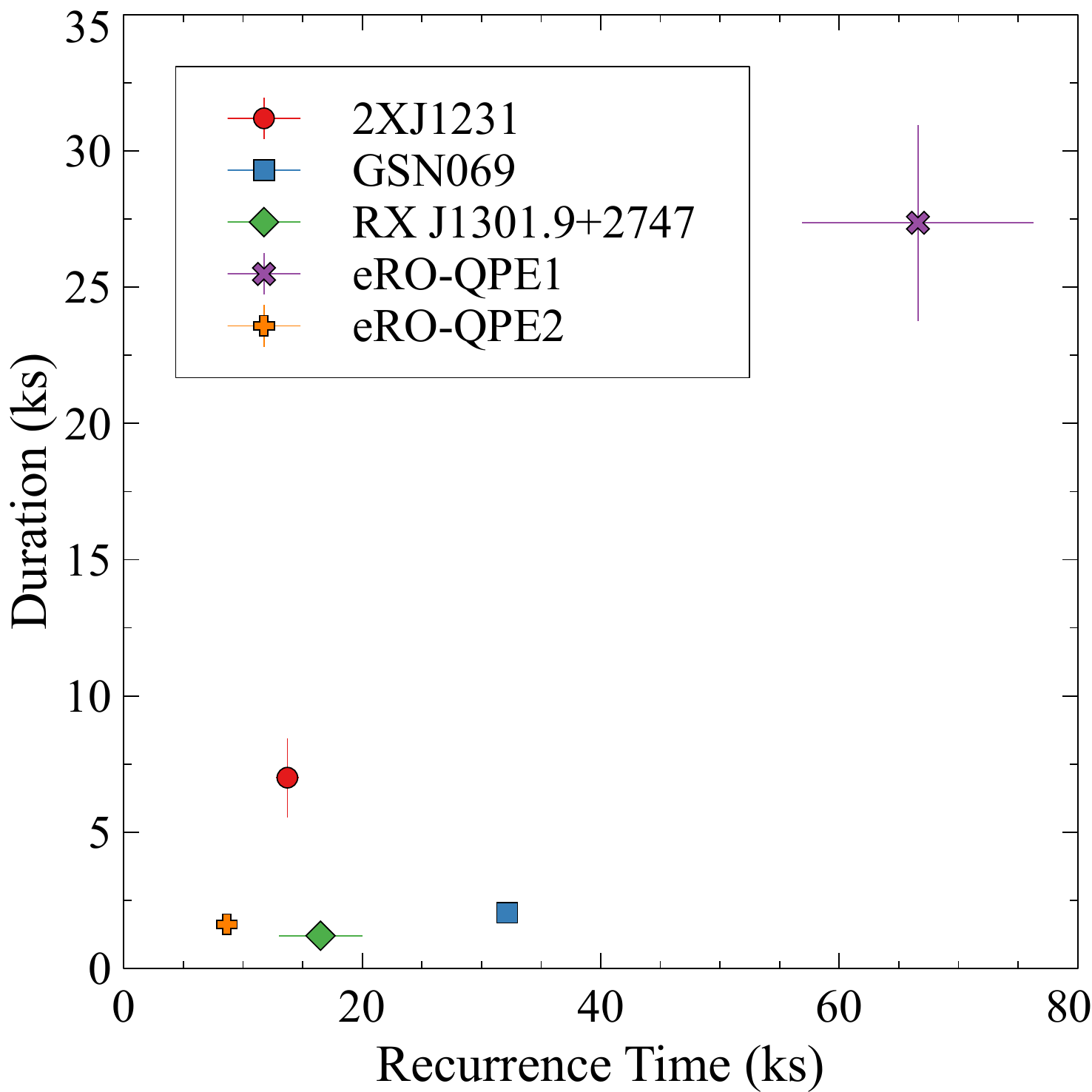}
    \caption{Distribution of $t_\text{rec}$ and $t_\text{dur}$ values as seen in 2XJ1231 and other QPE hosts. Values are taken from the literature and error bars are shown if available. For some objects the error bars may be smaller than the data point.}
    \label{fig:qpecand_recvsdur}
\end{figure}

In this analysis we have considered the use of the autocorrelation function in determining characteristic timescales of variability in an AGN, and have attempted to characterise the variability observed by fitting folded and unfolded lightcurves.

The autocorrelation function plots to the two observations of 2XJ1231 have presented two different characteristic timescales for variability: 13.52 ks for O1, and 14.35 ks for O2. The value of 13.71 ks as previously determined by \cite{Lin2012} is consistent within the errors determined by our analysis for both observations. Our value for the timescale for O1 is not, however, reconcilable with that found by \cite{Terashima2012} of 14.0 ks for the same observation. In both cases in the literature the timescale of variability was determined by means of a blind search in a comparison between folded lightcurves and a constant model. While it is possible, within the errors we have determined, for the timescale of variability to be constant there are indications that it is increasing as time passes, as was confirmed by the increased values for $t_\text{rec}$ in the sinusoidal and fixed Gaussian models in \ref{tab:lc_fits}. There were indications of an increasing recurrence time in early observations of GSN 069, as noted by \cite{Miniutti2019}, however further analysis of GSN 069 has revealed  a more complex picture, with subtly modulating recurrence times within observations \citep{Miniutti2022}. There is, however, a clearly different profile for the autocorrelation functions seen for the lightcurves of the two observations of 2XJ1231 in Figures \ref{fig:O1_full_lc_fit_and_acf} and \ref{fig:O2_full_lc_fit_and_acf} when compared with those for lightcurves which definitely contain QPEs, as in the case of Figure \ref{fig:ac_example}. The peaks seen in Figure \ref{fig:ac_example} are narrow, well separated and reach high levels of correlation, with the first two peaks in the autocorrelation plot of observation 0831790701 of GSN 069 being above 0.5. In contrast, the features seen in figures \ref{fig:O1_full_lc_fit_and_acf} and \ref{fig:O2_full_lc_fit_and_acf}  are much broader and of a lower magnitude. The value for the autocorrelation plot of O1 has a magnitude of 0.041 for $\tau$ = 13.52 ks, and the value of the autocorrelation for O2 is 0.037 where $\tau$ = 14.35 ks, as the first peaks of the fitted models to the autocorrelation functions, although the highest values in nearby points are over 0.05 and 0.04 for O1 and O2 respectively. 
The lower and broader peaks in the autocorrelation function confirm what is seen in the lightcurves; the variability seen is of a lower magnitude and with a higher proportion for the duty cycle ($t_\text{dur} / t_\text{rec}$) than that seen in observations where QPEs have previously been confirmed to be present. This also accounts for the strong anti-correlation seen at values of $(2n+1)t_\text{rec}/2$, which is not seen in the autocorrelation plots for observations with confirmed QPEs due to the longer periods of quiescence between eruptions in observations of other objects. For both observations, the peaks at $t_\text{rec}$ are significant at more than 5$\sigma$ (a confidence interval of 0.999999), with the significance levels for observations O1 and O2 being set at 0.0251 and 0.0211 respectively, calculated using the classical test for significance in the autocorrelation function as $(1+CI)/\sqrt{n}$ where CI is the confidence interval, and $n$ is the number of points in the lightcurve. This is complemented by the work of \cite{Lin2013}, where they determined using power spectra, and simulations of power law and broken power-law models, that a significant signal could be detected to $>5.1\sigma$. We also confirmed the significance to a high degree by simulating 1000 lightcurves for each observation using the method outlined by \cite{Done1992} with underlying red-noise power spectra, and computing the autocorrelations of these lightcurves. For O1 97.8$\%$ of the autocorrelation functions examined had peaks during the first correlated section after the first zero-crossing time lag at a lower level than that seen for O1, and for O2 the autocorrelation peak was higher than in 96.7$\%$ of the simulated lightcurves. The autocorrelation plots of O1 and O2 also contain strong periodic peaks at multiples of a given time lag which further point to a significant periodicity. A manual inspection of the $2.2\%$ of simulated lightcurves which had stronger individual autocorrelation peaks than that observed in O1 showed that none also had periodic signals, and the peaks were isolated. As such, we can report the significance of the four peaks seen in the autocorrelation plot of O1 to greater than 99.9$\%$ significance. A similar inspection of the $3.3\%$ of simulated lightcurves which had stronger individual autocorrelation peaks than that observed in O2 showed four with further peaks approximately located at time lags that were multiples of the first peak. As such, we can report the significance of the five peaks seen in the autocorrelation plot of O2 to greater than 99.6$\%$ significance.

The fitting to the folded lightcurves for both observations of 2XJ1231 showed only marginal differences in the goodness of fit when using a sinudoidal or Gaussian model, which became much greater when modelling the full lightcurves. For the folded lightcurves, the best fitting models were of Gaussian bright phases with a linearly decreasing and constant quiescent rate for observations O1 and O2 respectively. In the best Gaussian models the magnitudes of the bright phases, as the ratio of amplitude to quiescent rate ($A/x_\text{q}$), were 2.12 and 1.29 for observations O1 and O2, and the bright phases comprised 48.4$\%$ and 46.6$\%$ of the duty cycle, as the ratio of the duration and the recurrence time. These values for the duty cycle, while higher than any previously seen in QPE candidates, are not too dissimilar to one of those found by \cite{Arcodia2021} where one of the two galaxies exhibiting QPEs (eRO-QPE1) had a duty-cycle of 41$\%$. The values for the magnitudes of the bright phases which were determined in these full-band, folded, lightcurves are lower than any others seen in the literature. The reduced chi square values for the model fits to the two observations indicate that all of the models are significant at the 95$\%$ level for O2, but none are significant at even the 1$\%$ level for O1. Within the observations there is no significant difference in the $p$-values for the models tested and all give approximately the same quality of fit.

When fitting the models to the full lightcurves the goodness of fit was significantly reduced for the wave and Gaussian models when compared with the fitting to the folded lightcurves, and $\chi^2_\nu$ was not at a significant level even once the features of individual Gaussian bright phases were allowed to vary, although it was reduced significantly (reduced by 23.1$\%$ for O1 and by 15.3$\%$ for O2 on the next best performing models for approximately a 5$\%$ reduction in the degrees of freedom). While these reduced chi-square values are high they are not unreasonably higher than those which are achieved by fitting the variable Gaussian model with a linear baseline to observations which are known to contain eruptions: fit to observation 0831790701 of GSN 069 (Figure \ref{fig:gsn_fit}) with five eruptions had $\chi^2_\nu$=26.5 for 534 degrees of freedom; fit to observation 0851180501 of RX J1301.9+2747 (Figure \ref{fig:rx_fit}) with three eruptions had $\chi^2_\nu$=5.37 for 169 degrees of freedom. There were differences noted between the fits to the folded and the full lightcurves for $A$, $t_\text{dur}$ and $t_\text{rec}$. The Gaussian model when fit to the full lightcurves gave a slightly larger amplitude for bright phases than when applied to the folded observations. The amplitude of the bright phases for O1 and O2 (as $A/x_\text{q}$) were 2.60 and 3.46. The durations of bright phases were found to be shorter in O1 and longer in O2 than found with the folded fits. In order to provide a fair comparison between the sinusoidal model and another of equivalent complexity we then considered the fixed Gaussian model, and in the case of both observations the Gaussian model provided a better fit to the data than the sinusoidal model. The sinusoidal and fixed Gaussian models fit to O1 both underestimated the recurrence time against the results of the autocorrelation and those quoted in literature with the variable Gaussian model being the most accurate albeit with far larger uncertainties. For O2 all three models underestimated the recurrence time as per the results of the autocorrelation function, even beyond the uncertainties quoted, although the variable Gaussian model was consistent with the value for the recurrence time of 13.71 ks as quoted by \cite{Lin2013}. This could be indicative of a significantly varying recurrence time, as has already been seen in the eruptions of RX J1301.9+2747 where the recurrence time varies between $\sim$13.5 ks and $\sim$20.0 ks \citep{Giustini2020} and eRO-QPE1 which exhibits eruptions with a mean recurrence time of $\sim66.6\pm9.7$ ks. We then isolated each of the three factors which were allowed to vary in the variable Gaussian model in order to determine if one of them was more significant than the other two. For observation O1 (and O2) the values for $\chi^2_\nu$ were 4.68, 4.49 and 4.53 (4.61, 4.57 and 5.44) when A, $t_\text{dur}$ and $t_\text{rec}$ were the only features which were allowed to vary between bright phases in the same observation with a constant quiescent count rate. In both cases the most important feature to the quality of the fit was $t_\text{dur}$, but all three had a significant impact on the quality of fit for both observations.

Given that the profile of variability favours the Gaussian model over the sinusoidal model we now consider whether the variability seen fits the features of QPEs seen in other objects. The timescales of variability seen in 2XJ1231 are within the ranges of values seen in other QPE host objects. The average (and standard deviation on) $t_\text{rec}$ is $13.72 \pm 0.98$ ks, which is greater than that seen for eRO-QPE2, 8.64$\pm$0.30 ks \citep{Arcodia2021}, and more than 80$\%$ of the average recurrence time seen in RX J1301.9+2747, being 16.5$\pm$3.5 ks \citep{Giustini2020}, although it is considerably shorter than those seen in GSN 069 and eRO-QPE1, being 32.2$\pm$2.1 ks and 66.6$\pm$9.7 ks \citep{Miniutti2019,Arcodia2021}. The average of (and standard deviation on) the duration of bright phases seen in 2XJ1231, 7.00$\pm$1.45 ks, is also within the range of values for $t_\text{dur}$ seen in other objects. The largest duration for eruptions previously seen was in eRO-QPE1 at 27.4$\pm$3.6 ks \citep{Arcodia2021}, this object also having the longest recurrence time between durations. 2XJ1231 does display the largest duty cycle ($\Delta = t_\text{dur} / t_\text{rec}$) amongst any of the objects considered so far, with $\Delta = 48.4 \%$ in O1 and $\Delta = 46.6 \%$ in O2, the next largest being $\Delta \sim 41 \%$ for eRO-QPE1.

\begin{figure}
	\includegraphics[width=\linewidth]{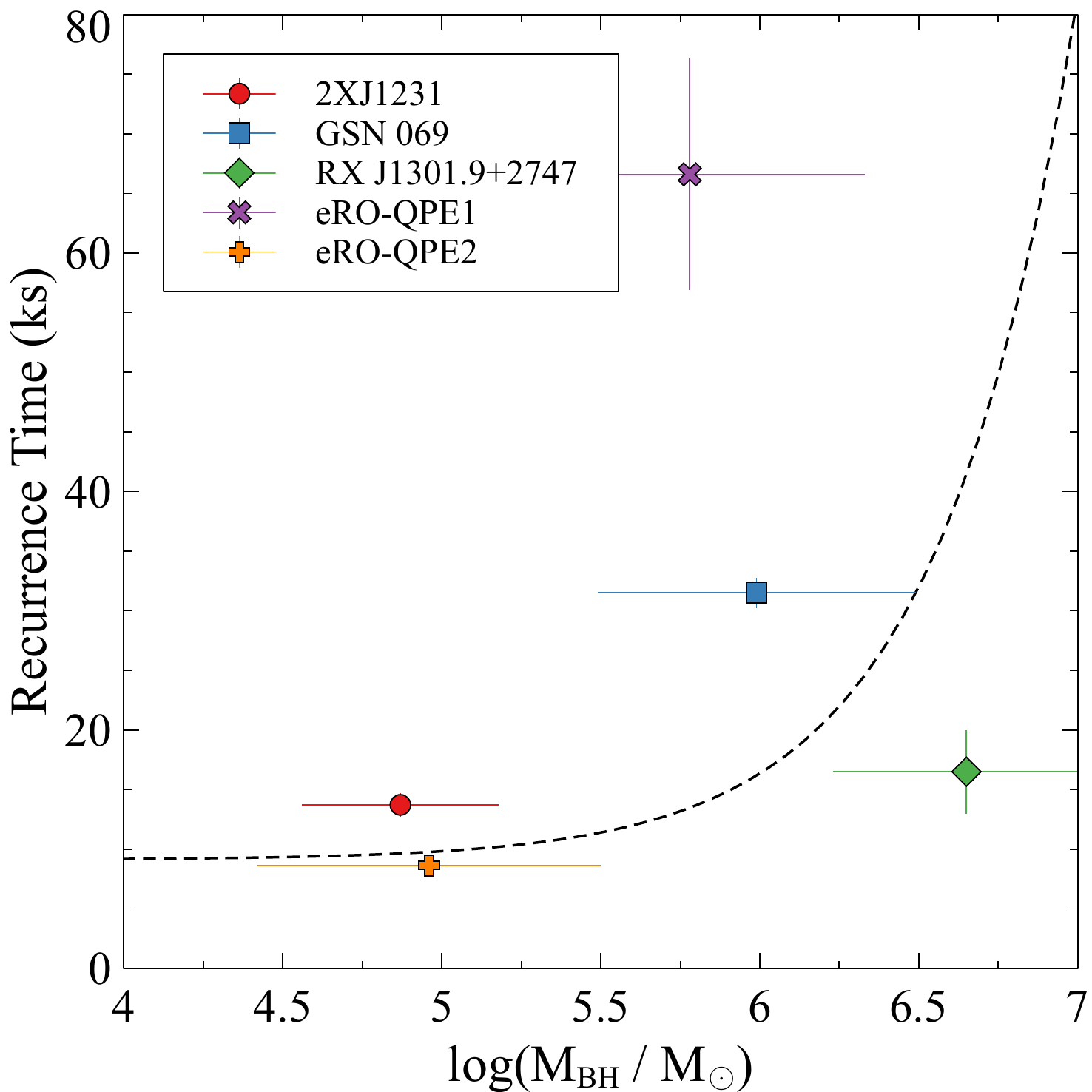}
    \caption{Distribution of $M_\text{BH}$ and $t_\text{rec}$ values as seen in 2XJ1231 and other QPE hosts. The dotted line indicates the best fitting linear relationship between $M_\text{BH}$ and $t_\text{rec}$.}
    \label{fig:qpecand_massvsrec}
\end{figure}

As has already been remarked \citep[][etc.]{Ho2012,Lin2013,Terashima2012}, 2XJ1231 has a very low black hole mass, $M_\text{BH}\lesssim10^5$. We therefore consider the possibility that the variability seen in 2XJ1231 may be the result of the same mechanism which drives QPEs in other objects, but around a less massive black hole. In Figure \ref{fig:qpecand_massvsrec} we show the distribution of $M_\text{BH}$ \citep{Wevers2022} and $t_\text{rec}$ \citep{Miniutti2019,Giustini2020,Arcodia2021} for QPE host AGN and 2XJ1231. We consider linear relationships between the mass and recurrence time under the assumption that timescales of variability could be dependent on black hole mass, in the case of thermal-viscous instabilities, or due to preferable radii around the central black holes which would scale with mass.

The best fitting relationship between $M_\text{BH}$ and $t_\text{rec}$ as determined by these five objects, see Fig.\ref{fig:qpecand_massvsrec}, was
\begin{equation}
    t_\text{rec} = \alpha + \beta \times \frac{M_\text{BH}}{M_\odot}
	\label{eq:mass_trec_rel}
\end{equation}
where $\alpha = 9.1 \pm 0.3$ ks and $\beta = (7.2 \pm 0.7) \times 10^{-6}$ ks/$M_\odot$ , although this has been heavily affected by the possible outlier RX J1301.9+2747, which deviates from the general trend for longer recurrence times being associated with higher black hole masses.  The long-term evolution of QPE timing properties will remain unclear, even if 2XJ1231 were to be considered a candidate. QPEs are clearly a transient phenomena in at least some of the sources already identified, with QPEs appearing and then disappearing in GSN 069, with a lifetime between 1 and 5.5 years \citep{Miniutti2022}. Conversely, if the eruptions detected in observations of RX J1301.9+2747 have continued between observations then the lifetime could be of the order of decades. The variability seen in 2XJ1231 'switched on' at some point between an observation in 2003 and those analysed above, and further observations would be needed to place limits on its lifetime.

\begin{figure}
	\includegraphics[width=\linewidth]{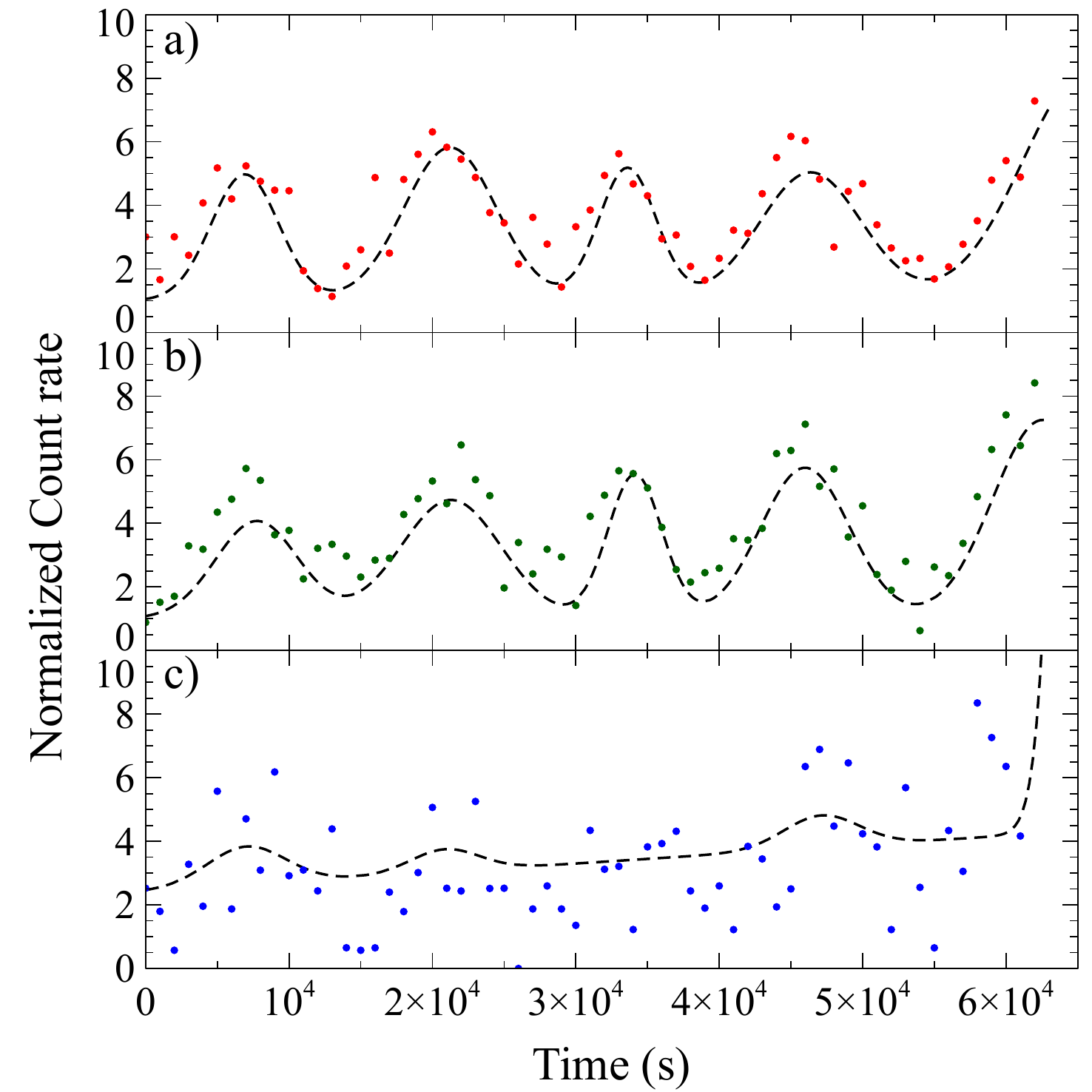}
    \caption{Lightcurves and associated Gaussian model fits for observation O1 of 2XJ1231 across three energy bands. Lightcurves and fits have been normalised by the quiescent count rates as determined by fitting and stated in Table \ref{tab:ebands_full}. a) Energy band 0.2-0.5keV. b) Energy band 0.5-1.0keV. c) Energy band 1.0-2.0keV.}
    \label{fig:o1_ebandfits}
\end{figure}

\begin{figure}
	\includegraphics[width=\linewidth]{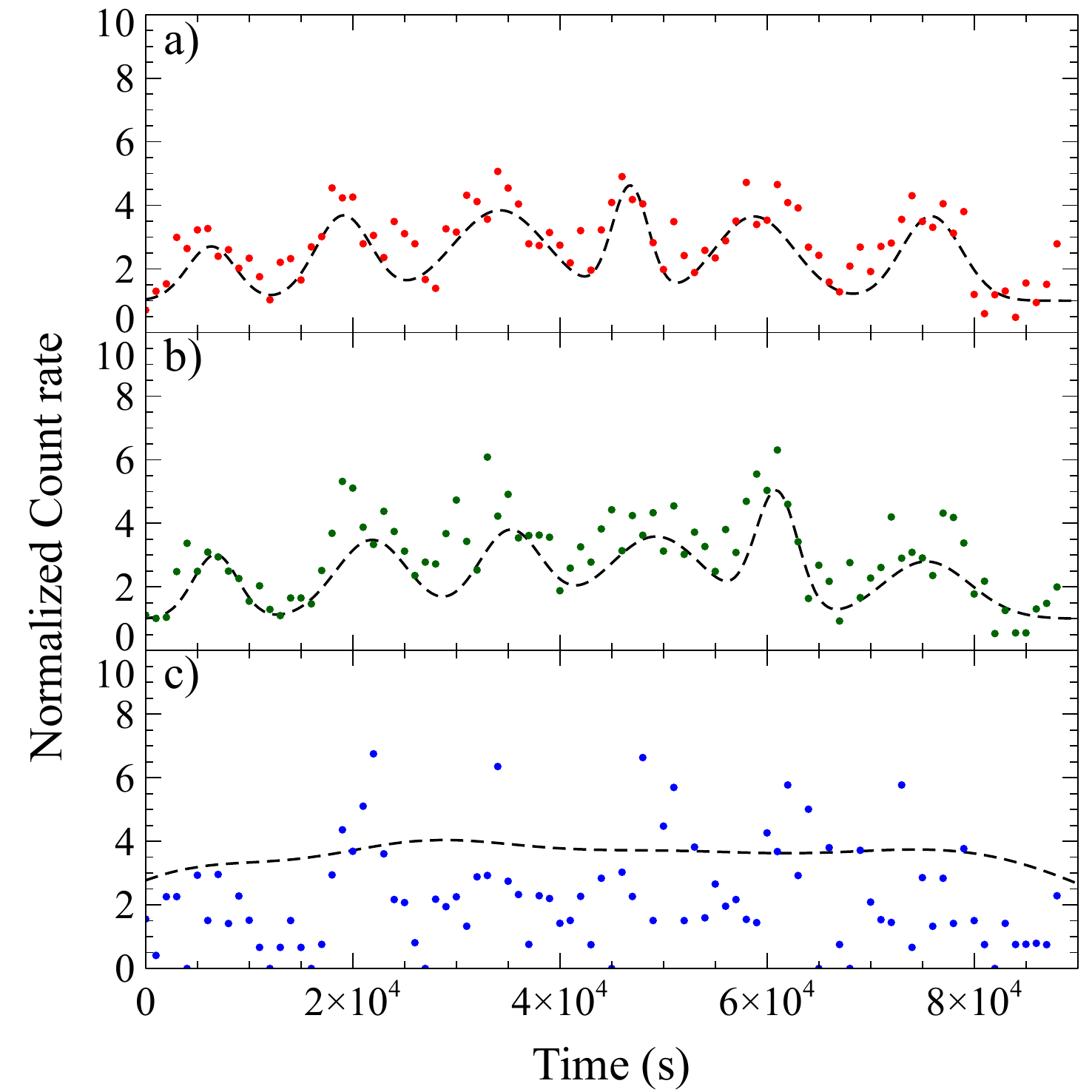}
    \caption{Lightcurves and associated Gaussian model fits for observation O2 of 2XJ1231 across three energy bands. Lightcurves and fits have been normalised by the quiescent count rates as determined by fitting and stated in Table \ref{tab:ebands_full}. a) Energy band 0.2-0.5keV. b) Energy band 0.5-1.0keV. c) Energy band 1.0-2.0keV.}
    \label{fig:o2_ebandfits}
\end{figure}

Several of the energy-dependent features of QPEs have been identified through the examination of eruptions in GSN 069 \citep{Miniutti2019} and RX J1301.9+2747 \citep{Giustini2020}, and are outlined above. We therefore consider the energy-dependent characteristics of the variability seen in 2XJ1231, using three energy bands 0.2--0.5~keV, 0.5--1.0~keV, and 1--2~keV. The lightcurves and best fitting models are shown in figures \ref{fig:o1_ebandfits} and \ref{fig:o2_ebandfits}. Details of the fits to the folded and full lightcurves are given in tables \ref{tab:ebands_folded} and \ref{tab:ebands_full} respectively. Given the super-soft nature of the spectrum, as previously identified by \cite{Lin2013} and \cite{Terashima2012} it is unsurprising that the variability in the highest energy band was dominated by noise (as was also noted in observations of GSN 069 above 2keV \cite{Miniutti2019} and above 1.3keV for RX J1301.9+2747 by \cite{Giustini2020}) and the fitter failed in both cases to satisfactorily fit the variable Gaussian model to the observed lightcurve. The lack of clear variability in the highest energy band observed does, however, continue to suggest that the variability seen in 2XJ1231 may indeed be indicative of the presence of QPEs in the lower energy ranges.  The effect of this band is to suppress the magnitude of variablility when viewed over a broader energy range. As such, the average amplitude of variability ($A/x_\text{q}$) found using the variable Gaussian model in the 0.2-1.0keV energy band is 2.86$\pm$0.45 for O1 and 8.56$\pm$1.14 for O2.

\begin{table}
	\centering
	\caption{Parameters obtained from Gaussian bright phase fits to the folded lightcurves of observations O1 and O2 of 2XJ1231 in three energy bands. The model used in fitting used a constant quiescent count rate. Parameters are as described in section \ref{sec:models}.}    
	\label{tab:ebands_folded}
	\begin{tabular}{lcccr} %
		\hline
		Obs. & Energy (keV) & $A/x_\text{q}$ & $t_\text{0}$ (s) & $t_\text{dur}$ (s)\\
		\hline
		O1 & 0.2-0.5 & 9.70$\pm$5.76 & 6770$\pm$70 & 7790$\pm$220 \\
		O1 & 0.5-1.0 & 1.65$\pm$0.17 & 6930$\pm$10 & 5180$\pm$10 \\
		O1 & 1.0-2.0 & 0.301$\pm$0.204 & 8520$\pm$7040 & 3020$\pm$40 \\
		O2 & 0.2-0.5 & 1.57$\pm$0.20 & 4570$\pm$140 & 5920$\pm$390 \\
		O2 & 0.5-1.0 & 14.7$\pm$18.8 & 5040$\pm$120 & 10000$\pm$100 \\
		O2 & 1.0-2.0 & 1.53$\pm$2.92 & 6230$\pm$4070 & 5000$\pm$10 \\
		\hline
	\end{tabular}
\end{table}

The fits to the folded lightcurves do, at times, show the energy-dependent characteristics of QPEs which have been seen in other objects. In observation O1 the durations of bright phases appear to decrease in time as the energy band increases, and is also seen in the highest and lowest energy bands in O2, with the 0.5-1.0~keV band being an exception. We also observe a much larger amplitude for bright phases in O2 in the 0.5-1.0~keV band as opposed to the bright phases in the 0.2-0.5~keV band. Contrary to observations of GSN 069 and RX J1301.9+2747, however, the bright phases appear to occur later with increasing energy, as opposed to earlier \citep{Miniutti2019,Giustini2020}. There is a similarly unclear pattern of results when considering the fits to the full lightcurves, with the amplitude decreasing with energy in O1 and the reverse seen in O2. Additionally, $t_\text{dur}$ is lowest in the 0.5-1.0keV band in O1, but the durations of bright phases increases with energy in O2. Such energy-dependent behaviour is contrary to that which is seen in other QPE sources \citep{Miniutti2019,Giustini2020}.

The highly energy dependent amplitudes of the eruptions seen in 2XJ1231 are of a generally lower size than those seen in other QPE host objects, although this is very much energy dependent. The amplitudes of eruptions vary between $\sim$2 and $\sim$100 in the 0.2-0.3~keV and 0.6-0.8~keV bands in GSN 069, between $\sim$2 and $\sim$100 in the 0.2-0.3~keV and 1.0-1.3~keV bands in RX J1301.9+2747, and the ratio of peak to quiescent 0.001-100~keV disk luminosities in eRO-QPE1 and eRO-QPE2 are 294 and 11 respectively. The amplitudes of the eruptions reaching as much as a factor of ten in 2XJ1231 would therefore widen the parameter space for known QPE sources should this indeed be considered as the fifth QPE host object.

\begin{table}
	\centering
	\caption{Parameters obtained from variable Gaussian fits to the full lightcurves of observations O1 and O2 of 2XJ1231 in three energy bands. Values stated for A/$x_\text{q}$ and $t_\text{dur}$ are mean and standard deviations across the bright phases in the fitted model. Parameters are as described in section \ref{sec:models}.}
	\label{tab:ebands_full}
	\begin{tabular}{lcccr} %
		\hline
		Obs. & Energy (keV) & $x_\text{q}$ & $A/x_\text{q}$ & $t_\text{dur}$ (s)\\
		\hline
		O1 & 0.2-0.5 & 0.016$\pm$0.005 & 4.87$\pm$1.26 & 6720$\pm$520 \\
		O1 & 0.5-1.0 & 0.012$\pm$0.002 & 4.47$\pm$1.08 & 6710$\pm$10 \\
		O1 & 1.0-2.0 & 0.011$\pm$0.001 & 0.84$\pm$0.51 & 7630$\pm$150 \\
		O2 & 0.2-0.5 & 0.017$\pm$0.001 & 2.68$\pm$0.53 & 4440$\pm$480 \\
		O2 & 0.5-1.0 & 0.008$\pm$0.001 & 4.0$\pm$0.9 & 4980$\pm$180 \\
		O2 & 1.0-2.0 & 0.010$\pm$0.001 & 0.10$\pm$0.16 & 6230$\pm$400 \\
		\hline
	\end{tabular}
\end{table}

The temperature of the disk, as determined by the spectral fitting reported in section \ref{subsec:energyspec}, does not change within errors, neither across observations nor during the different bright and low phases. Instead, we observe a significant change in the normalisation, indicating that the variability may be due to a change in the area of the radiating region, rather than in the temperature of the radiative material. This is in contrast with spectral fitting to other QPE objects, which show transitions from cooler states during quiescence to hotter states during eruptions \citep{Miniutti2019,Giustini2020,Chakraborty2021,Arcodia2021}, with typical temperature differences between quiescent and eruptive states being $\gtrsim 50$~eV. The temperature is closer to that associated with the eruptive phases of QPE hosts, and so the absence of temperature changes may be caused by the lack of real quiescent states in 2XJ1231 caused by broader, overlapping bright phases. It is notable that this temperature is also comparable to that of the soft excess seen in AGN.

Ultimately, the analysis we can perform is limited by the brightness of the source and the sensitivity of detectors and observations. The available observations are less than the limit of what is possible with regards to the exposure time and \emph{XMM} orbit restrictions, but given that we have already observed a considerable number of variability cycles we should focus on obtaining pointed observations, rather than those where the source is a long way off-axis, which might improve the data and allow MOS1 to be used. Given the super-soft nature of the source spectrum it is likely that we are approaching the limits of what can be learned about 2XJ1231 using \emph{XMM} unless it becomes significantly brighter, which would seem unlikely given that follow-up \emph{Swift} and \emph{Chandra} observations \citep{Lin2017} indicate that the source is becoming fainter. If this decrease in luminosity is due to the long-term evolution of a tidal disruption event (TDE) it could indicate that the variability seen in 2XJ1231 has a similar cause to that seen in GSN 069 and XMMSL1 J024916.6-041244, both of which are considered to contain QPEs, and which are both strong TDE candidates \citep{Chakraborty2021,Miniutti2022}. If this is the case we might expect to see a re-brightening as the variability seen in O1 and O2 disappears as has been observed in GSN 069 \citep{Miniutti2022}.

Given the ambiguous nature of these results we then conducted a manual, targeted search for other AGN which might contain QPEs in the \emph{XMM} archive. To determine a sample of AGN to look at in greater detail we used some of the criteria which were used to identify RX J1301.9+2747 and 2XJ1231 as possible QPE candidates. To find AGN which meet these criteria we used two freely-available enhanced catalogues CHANSNGCAT\footnote{\url{https://heasarc.gsfc.nasa.gov/W3Browse/all/chansngcat.html}} \citep{She2017}, and Optical Emission Line Properties and Black Hole Mass Estimates for SPIDERS DR14 AGN\footnote{\url{https://www.sdss.org/dr15/data_access/value-added-catalogs/?vac_id=optical-emission-line-properties-and-black-hole-mass-estimates-for-spiders-dr14-agn}} \citep{Coffey2019}. Making an initial selection based only on $M_\text{BH}$, as determined using the $M_\text{BH}-\sigma$ relationship for CHANSNGCAT and using spectral lines for SPIDERS DR14 AGN, gave a very restrictive sample, with only 17 AGN with $\log(M_\text{BH} / M_\odot) \lesssim 6$ identified in the CHANSNGCAT catalogue, and none below that threshold in the SPIDERS catalogue (with $\log(M_\text{BH} / M_\odot)=6.18$ being the lowest). From the 17 AGN identified using CHANSNGCAT we found 11 objects which had a combined 51 observations in the \emph{XMM} archive, details of which are presented in Table \ref{tab:cand_obs} in Appendix \ref{appendix:A}. The observations were processed in the same manner as for the observations of 2XJ1231 analysed above, and lightcurves from the EPIC pn instrument for events from 0.2-2.0keV and associated autocorrelation plots were produced. In none of the lightcurve or autocorrelation plots for any of the 51 observations of these 11 objects were there any features which indicated the possible existence of QPEs. Although this search was only performed on a very select sample it suggests that variability of the kind seen in GSN 069, RX J1301.9+2747, eRO-QPE1, eRO-QPE2, and 2XJ1231 is not a given feature of all lower-mass AGNs and may be even more unusual than previously thought.

\section{Conclusions}
\label{sec:conc}

The variability seen in the low-mass, super-soft AGN 2XMM J123103.2+110648 remains somewhat ambiguous in nature. We have successfully applied the autocorrelation function to identify characteristic timescales for the variablity, and found values which are consistent with those determined by other means as quoted in the literature. The profiles of the autocorrelation plots were different in appearance to those seen for observations with confirmed QPEs, being less pronounced, but the features seen were still statistically significant at 5$\sigma$, with $t_\text{rec}$ = 13.52 ks and 14.35 ks for O1 and O2 respectively. This analysis has shown that the autocorrelation function can be applied in finding QPE-like behaviour, and as such it should be considered as a viable tool in further attempts to find more objects which could contain QPEs.

\begin{figure}
	\includegraphics[width=\linewidth]{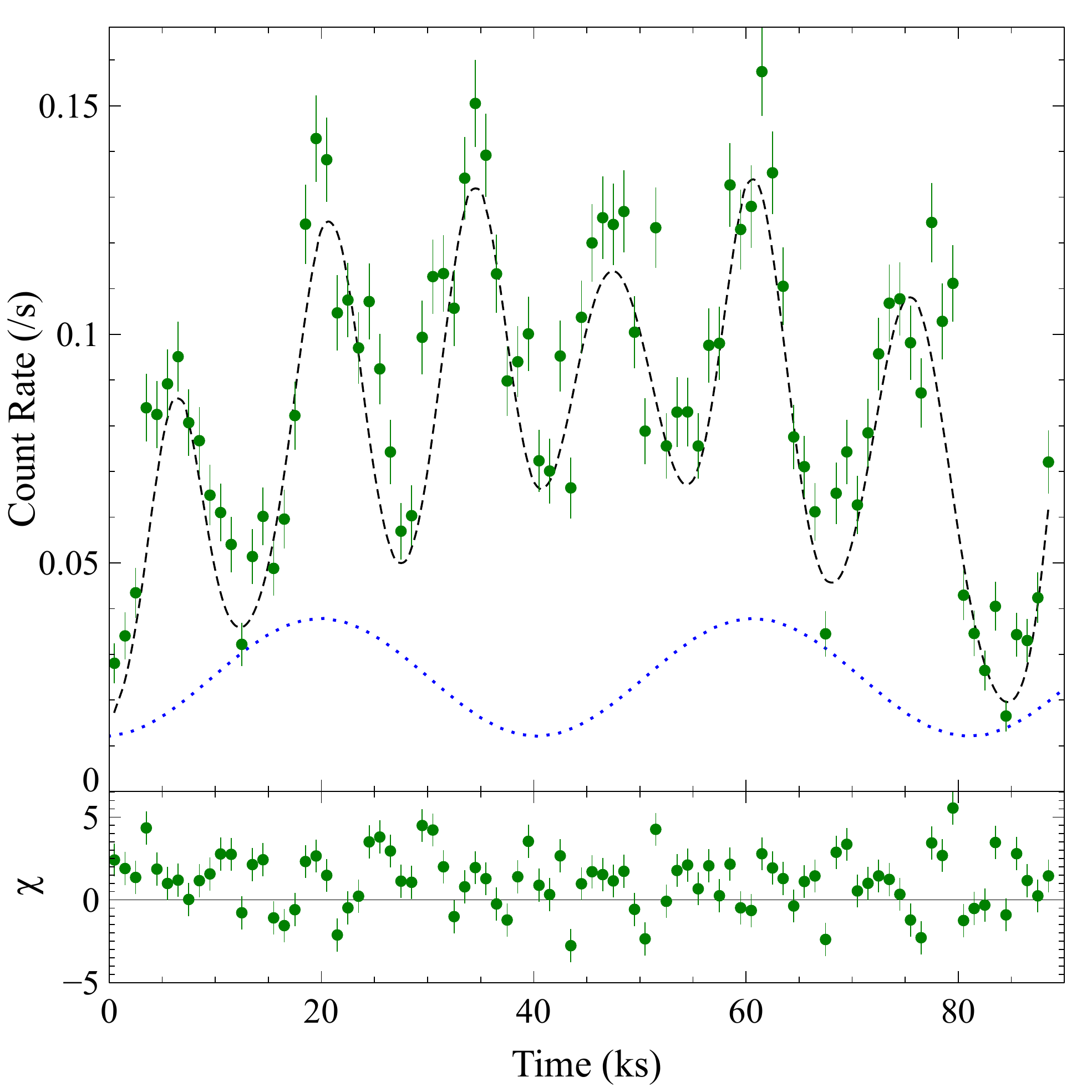}
    \caption{Lightcurve and associated Gaussian eruption with oscillating baseline model fit, for observation O2 of 2XJ1231. The black dashed line shows the overall model fit, and the blue dotted line shows the component of that model which constitutes the oscillating baseline. The lower panel shows the deviations of individual points from the fitted model.}
    \label{fig:O2_oscillatingbaseline_fit}
\end{figure}

We have considered the profile of the variability seen using a number of measures and find that they point towards the variability in 2XJ1231 being a QPE in nature. We determined from the best fitting variable QPE models that the amplitudes (as $A$/$x_\text{q}$), recurrence times and durations of eruptions in a 0.2--2.0~keV energy range were 2.60 (3.46), 13.7 ks (13.7 ks), and 6.17 ks (7.69 ks) respectively for O1 (and O2). The profile of the variability is best described by a series of Gaussian eruptions, and although the models provide a reasonable fit, they are too rudimentary to fully describe the behaviour. We have observed that the baseline rate is likely more complex than a linear trend, and indeed the fit to observation 0831790701 of GSN 069 (see Figure \ref{fig:gsn_fit}) is suggestive of an oscillating quiescent count rate with there also being an indication of a QPO during the quiescent periods of RX J1301.9+2747 \citep{Song2020}. A brief examination of the possibility of an oscillating baseline count rate for the Gaussian models significantly improved the fit to O2, however there was no improvement in the fit quality for O1. For O1, the inclusion of an oscillating baseline resulted in $\chi^2_\nu=3.82$ for 234 degrees of freedom, while for O2 the fit improved, with $\chi^2_\nu=4.17$ for 334 degrees of freedom. This fit to O2 is shown in Figure \ref{fig:O2_oscillatingbaseline_fit}. 2XJ1231 would be the lowest mass AGN should it be the fifth member of the newly discovered class of QPE host galaxies. 

We have purposely taken an agnostic approach to this profiling with regards to the cause or method by which eruptions could be created in order to not prejudice the identification of the features which we have identified. In now considering some of the models which have previously been proposed by others to explain QPE phenomena we find that:
\begin{itemize}
    \item While the case against gravitational lensing, as described by \cite{Ingram2021}, for the mechanism in QPE sources is strong in light of the asymmetry seen in the eruptions of eRO-QPE1 \citep{Arcodia2022}, we can make no similar claim for the variability in 2XJ1231. It does, however, still appear to be an unlikely mechanism, as with GSN 069 and RX J1301.9+2747, as the black hole mass being as low as $10^5 M_{\odot}$ would produce eruptions which are either not of a large enough amplitude or not long enough in duration without the inclusion of a complex geometry for the emitting regions.
    \item An accretion flow instability, following the same methodology as applied by \citet{Arcodia2021} and outlined in \citet{Grzedzielski2017}, for the observation O2, with amplitude, as the ratio of the unabsorbed 0.5--2.0 keV luminosities in the high and low flux states \citep{Terashima2012} as a proxy for the bolometric luminosities, $A=3.29$, duty cycle $\Delta=0.484$, and recurrence time $t_\text{rec}=13.52$ ks gives a value for the viscosity parameter of $\alpha=0.27$ for $M_\text{BH}=10^5 M_\odot$. While this value for $\alpha$ is not unfeasibly high it would increase with a less conservative estimate for the amplitude of the eruptions.
    \item Accretion from a compact companion object, as described by \citet{King2020}, would imply that the orbiting companion is a white dwarf of mass $0.183 M_\odot$ in an highly eccentric orbit with $e=0.947$. The pericentre for such an orbit would be at $6.3 r_g$. This mechanism would raise further questions as to the high duty-cycle seen in the variability of 2XJ1231. If we were to consider the possibility of a less eccentrically orbiting companion, which would be accreting for a greater proportion of its orbit, rather than just at pericentre passage, we might expect to see a smoother transition between the high and low states, and a greater duty cycle such as is seen in 2XJ1231. Whether this could be extended further to explain the behaviour of QPO hosts is unclear.
\end{itemize}

We have also examined a wider population of other low-mass AGN, with eleven objects identified as having $\log(M_\text{BH} / M_\odot) \lesssim 6$, and examined 47 \emph{XMM} observations of these objects. We found no evidence of QPE-like behaviour in examinations of the lightcurves and autocorrelation plots of these observations over a range of time-binning space.

The immediate aims for further work with 2XJ1231 should include
\begin{itemize}
    \item Obtaining more targeted observations in order to provide better quality lightcurves with higher SNR.
    \item Applying more complex models to the variability seen to better profile the behaviour of 2XJ1231.
\end{itemize}

In moving forward we also need to consider how to appropriately distinguish between quasi-periodic oscillations and eruptions, or indeed whether they are in fact different manifestations of the same phenomenon. Does the variability seen in 2XJ1231 represent an intermediate point on the QPO/QPE continuum? We have considered the profiling of lightcurves as a possible mechanism, but other parametric measures should also be considered if a distinction between the two pheomena is necessary. Continued monitoring of the other QPE host objects to determine how their QPEs evolve over time will also help to build a fuller understanding of quasi-periodic eruptions, how they fit into our understanding of AGN, and of how 2XJ1231 fits into our understanding of quasi-periodic eruptions.

\section*{Acknowledgements}

All figures in this paper were created using Veusz. This work is supported by the UKRI AIMLAC CDT, funded by grant EP/S023992/1. We sincerely thank the anonymous referee for their thoughtful and constructive report which improved the quality of this paper.

\section*{Data $\&$ Code Availability}

All data used in this analysis is freely available in the \emph{XMM} Newton Science Archive (\url{http://nxsa.esac.esa.int/nxsa-web/#home}). This analysis used the \texttt{stingray} module for \texttt{python} version 0.3 and all code and data analysis products are available on GitHub upon request. For the purpose of open access, the author has applied a Creative Commons Attribution (CC BY) licence [where permitted by UKRI, ‘Open Government Licence’ or ‘Creative Commons Attribution No-derivatives (CC BY-ND) licence’ may be stated instead] to any Author Accepted Manuscript version arising




\bibliographystyle{mnras}
\typeout{}
\bibliography{qpe_refs} 




\appendix

\section{Appendix A - QPE Candidate search}
\label{appendix:A}

Table \ref{tab:cand_obs} contains a list of all \emph{XMM} observations of QPE candidate galaxies which were processed and analysed for evidence of QPEs. In all cases the pn lightcurves were binned at multiples of 10s for all values between 10s and 500s, to account for shot-noise in lower count rate observations, and associated autocorrelation plots were created at all time bin values. In none of the observations listed was any QPE-like behaviour observed.

\begin{table}
	\centering
	\caption{Details of observations of low-mass AGN which were identified as part of a manual search for QPE candidates. Values for $\log(M_\text{BH} / M_\odot)$ are as per CHANSNGCAT.}
	\label{tab:cand_obs}
	\begin{tabular}{cccr} %
		\hline
		AGN Name & $\log(M_\text{BH} / M_\odot)$ & OBSID & Exposure (ks) \\
		\hline
		NGC1331 & 5.48 & 0304190101 & 65.9 \\
		NGC3367 & 5.64 & 0551450101 & 31.4 \\
		NGC3599 & 5.85 & 0411980101 & 6.9 \\
		 -- & --  & 0556090101 & 43.6 \\
		NGC4467 & 5.86 & 0112550601 & 24.6 \\
		 -- & -- & 0200130101 & 111.0 \\
		 -- & -- & 0510011501 & 10.1 \\
		 -- & -- & 0761630101 & 118.0 \\
		 -- & -- & 0761630201 & 118.0 \\
		 -- & -- & 0761630301 & 117.0 \\
		NGC4476 & 5.58 & 0114120101 & 60.1 \\
		 -- & -- & 0200920101 & 109.3 \\
		 -- & -- & 0551870401 & 21.6 \\
		 -- & -- & 0551870601 & 21.3 \\
		 -- & -- & 0603260201 & 17.9 \\
		 -- & -- & 0803670501 & 132.0 \\
		 -- & -- & 0803670601 & 65.0 \\
		 -- & -- & 0803671001 & 63.0 \\
		 -- & -- & 0803671101 & 131.9 \\
		NGC4559 & 5.14 & 0152170501 & 42.2 \\
		 -- & -- & 0842340201 & 75.4 \\
		NGC4654 & 5.07 & 0651790201 & 28.9 \\
		NGC5273 & 5.97 & 0112551701 & 17.1 \\
		 -- & -- & 0805080401 & 110.9 \\
		 -- & -- & 0805080501 & 28.0 \\
		NGC6946 & 5.43 & 0093641201 & 8.5 \\
		 -- & -- & 0093641501 & 8.6 \\
		 -- & -- & 0093641601 & 10.1 \\
		 -- & -- & 0093641701 & 11.3 \\
		 -- & -- & 0200670101 & 16.4 \\
		 -- & -- & 0200670201 & 14.4 \\
		 -- & -- & 0200670301 & 15.6 \\
		 -- & -- & 0200670401 & 21.2 \\
		 -- & -- & 0401360101 & 20.9 \\
		 -- & -- & 0401360201 & 24.4 \\
		 -- & -- & 0401360301 & 24.4 \\
		 -- & -- & 0500730101 & 31.9 \\
		 -- & -- & 0500730201 & 37.3 \\
		 -- & -- & 0691570101 & 119.3 \\
		 -- & -- & 0794581201 & 50.0 \\
		 -- & -- & 0870830101 & 17.9 \\
		 -- & -- & 0870830201 & 17.7 \\
		 -- & -- & 0870830301 & 16.0 \\
		 -- & -- & 0870830401 & 17.8 \\
		NGC7314 & 5.59 & 0111790101 & 44.7 \\
		 -- & -- & 0311190101 & 83.9 \\
		 -- & -- & 0725200101 & 140.5 \\
		 -- & -- & 0725200301 & 132.1 \\
		 -- & -- & 0790650101 & 65.0 \\
		NGC925 & 6.0 & 0784510301 & 50.0 \\
		 -- & -- & 0862760201 & 42.0 \\
		\hline
	\end{tabular}
\end{table}

\section{QPE Model Fits to GSN 069 and RX J1301.9+2747}
\label{appendix:B}

Figures showing the QPE model fits, with variable features and linearly increasing (zero within errors) baseline count rate for observations 0831790701 of GSN 069 and 0851180501 of RX J1301.9+2747. Lightcurves were binned at a rate of 250s for fitting and are presented with time bins of 1 ks.

\begin{figure}
	\includegraphics[width=\linewidth]{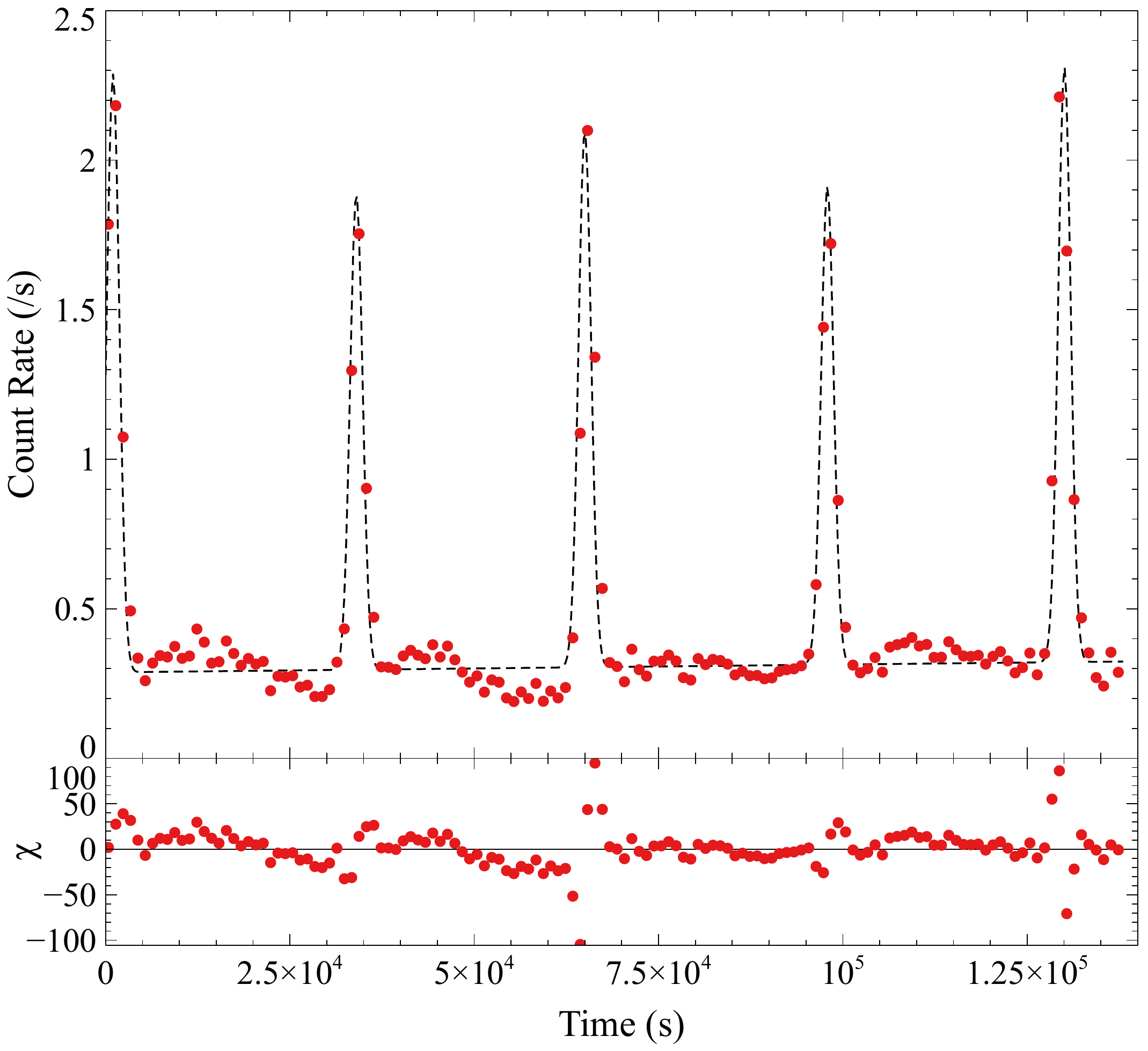}
    \caption{Lightcurve and associated Gaussian eruption model fit for observation 0831790701 of GSN 069. The QPE model used five eruptions and achieved a fit of $\chi^2_\nu$=26.5 for 534 degrees of freedom.}
    \label{fig:gsn_fit}
\end{figure}

\begin{figure}
	\includegraphics[width=\linewidth]{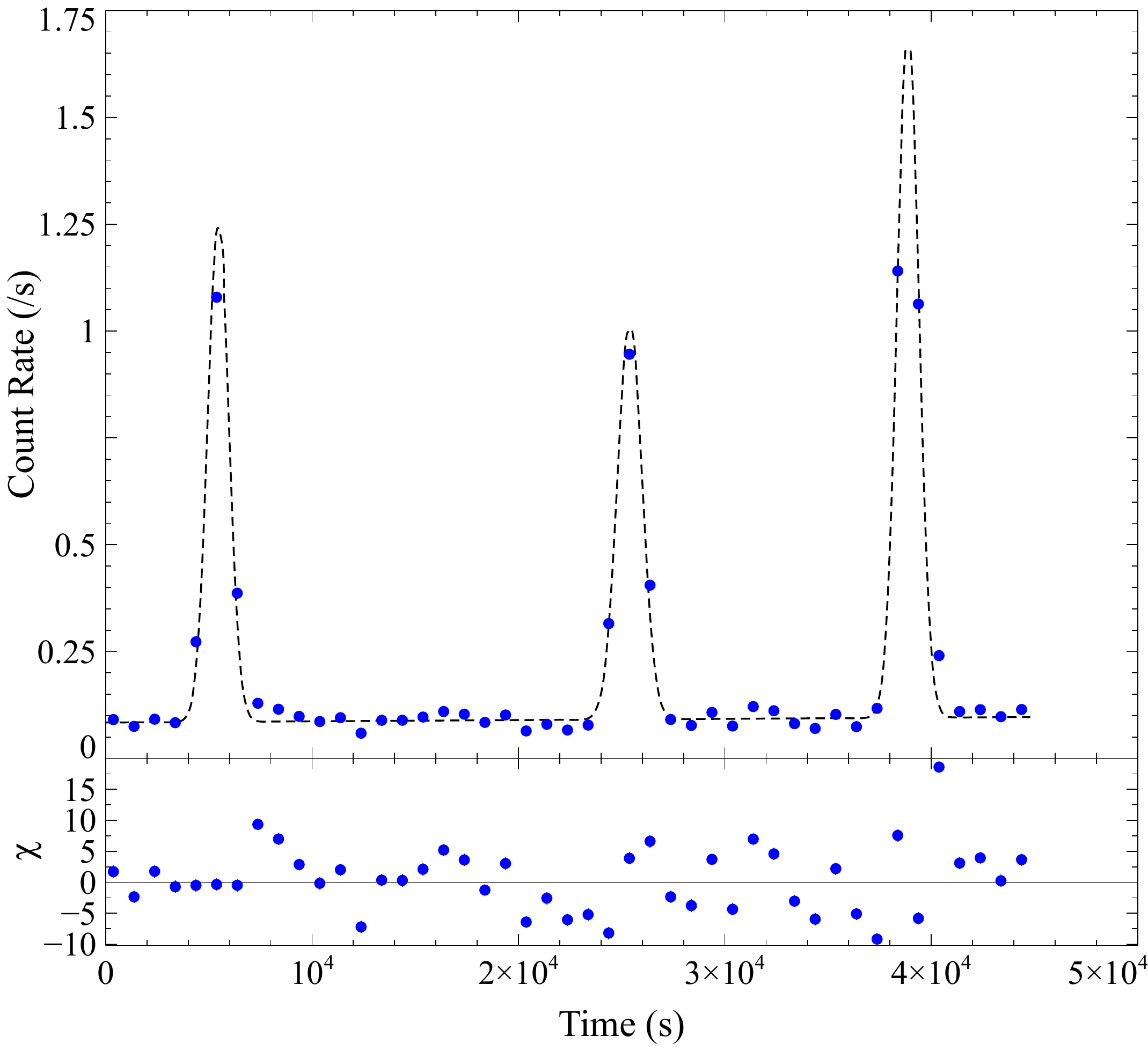}
    \caption{Lightcurve and associated Gaussian eruption model fit for observation 0851180501 of RX J1301.9+2747. The QPE model used three eruptions and achieved a fit of $\chi^2_\nu$=5.37 for 169 degrees of freedom.}
    \label{fig:rx_fit}
\end{figure}


\bsp	
\label{lastpage}
\end{document}